\documentclass[a4paper,dvips,12pt]{article}

\usepackage{epsf}
\usepackage[dvips]{graphicx}
\usepackage{tabularx}
\usepackage{latexsym}
\usepackage{amsmath,amssymb,exscale}
\usepackage{array,multicol}
\numberwithin{equation}{section}

\def\id{\ 1 \! \! \! \! 1}
\def\Tr{{\rm Tr}}
\newcommand{\be}{\begin{equation}}
\newcommand{\ee}{\end{equation}}
\newcommand{\ba}{\begin{eqnarray}}
\newcommand{\ea}{\end{eqnarray}}
\def\simlt{\mathrel{\lower2.5pt\vbox{\lineskip=0pt\baselineskip=0pt
             \hbox{$<$}\hbox{$\sim$}}}}
\def\simgt{\mathrel{\lower2.5pt\vbox{\lineskip=0pt\baselineskip=0pt
             \hbox{$>$}\hbox{$\sim$}}}}

\title{
\vspace*{-0.8cm}
\begin{flushright}
\normalsize{CERN--PH--TH/2005-136\\
\normalsize{IC/2005/042; LMU-ASC 55/05}\\
\texttt{hep-th/0507244}}\\
\end{flushright}
\vspace{.5cm}
\Large\textbf{Open string topological amplitudes and gaugino masses}
\author{\large
{\bf I.~Antoniadis$^1$\footnote{On leave from CPHT
(UMR CNRS 7644) Ecole Polytechnique, F-91128 Palaiseau},
K.~S.~Narain$^{2}$, T.~R.~Taylor$^{3}$}\\ \\
\emph{$^1$Department of Physics, CERN - Theory Division}\\
\emph{CH--1211 Geneva 23, Switzerland}\\
\emph{$^2$International Centre for Theoretical Physics}\\
\emph{I--34100 Trieste, Italy}\\
\emph{$^3$Department of Physics, Northeastern University}\\
\emph{Boston, MA 02115, U.S.A.}}}
\date{}
\begin{document}
\maketitle

\thispagestyle{empty}
\vspace*{-.5cm}

\begin{abstract}
We discuss the moduli-dependent
couplings of the higher derivative F-terms $(\Tr W^2)^{h-1}$, where
$W$ is the gauge $N=1$ chiral superfield. They are
determined by the genus zero topological partition function $F^{(0,h)}$,
on a world-sheet with $h$ boundaries. By string duality, these
terms are also related to heterotic topological amplitudes studied in
the past, with the topological twist applied only in the left-moving
supersymmetric sector of the internal $N=(2,0)$ superconformal field theory.
The holomorphic anomaly of these couplings relates them to
terms of the form $\Pi^n({\rm Tr}W^2)^{h-2}$, where $\Pi$'s represent
chiral projections of non-holomorphic functions of chiral superfields.
An important property of these couplings is that they violate R-symmetry
for $h\ge 3$. As a result, once supersymmetry is broken by
D-term expectation values, $(\Tr W^2)^2$ generates gaugino masses
that can be hierarchically smaller than the
scalar masses, behaving as $m_{1/2}\sim m_0^4$ in
string units. Similarly, $\Pi{\rm Tr}W^2$ generates Dirac masses for
non-chiral brane fermions, of the same order of magnitude.
This mechanism can be used for instance
to obtain fermion masses at the TeV scale for scalar masses as high as
$m_0\sim{\cal O}(10^{13})$ GeV. We present explicit examples
in toroidal string compactifications with intersecting D-branes.
\end{abstract}
\date

\newpage

\section{Introduction} \label{Intro}

Possible relation between string theory and topological field theories
may have profound origin and help in understanding the non-perturbative
nature of the underlying fundamental theory. A concrete example is the
topological partition function of the twisted Calabi-Yau sigma-model,
$F^{(g,h)}$, defined on a general Riemann surface of genus $g$ having
$h$ boundaries~\cite{bcov}. In the absence of holes, it is known that $F^{(g,0)}$
describes a sequence of higher derivative F-terms, ${\cal W}^{2g}$, in the
effective four-dimensional (4d) $N=2$ supergravity, obtained upon
compactification of Type II superstring on the corresponding Calabi-Yau
manifold~\cite{agnt,bcov}. Here, ${\cal W}$ is the chiral superfield of the $N=2$ gravity
multiplet. Although $F^{(g,0)}$ is expected to be a holomorphic function
of the chiral moduli, there is an anomaly which is captured by a set of recursion
relations~\cite{bcov}. This non-holomorphicity is induced by an anomaly of the
BRST current in the topological theory, while from the space-time point
of view, it is due to the propagation of massless states that lead to
non-localities in the effective action~\cite{agnt}.

In this work, we perform a similar study of the genus-0 series $F^{(0,h)}$.
In the presence of boundaries, $h\ne 0$, the topological twist involves
D-branes, and thus supersymmetry is broken to $N=1$.
The partition function $F^{(0,h)}$ describes a
sequence of higher derivative F-terms, $(\Tr W^2)^{h-1}$, where $W$
is the chiral $N=1$ gauge superfield \cite{bcov,ov,bov}. They give rise to amplitudes
involving two gauge fields and $2(h-2)$ gauginos. In particular, we rederive the relation between the scattering amplitudes and the topological partition function by using the Neveu-Schwarz-Ramond formalism, similarly to Ref.\cite{agnt}.
The non-trivial
property of this result is that in the physical amplitudes, besides
the cancellation of the string oscillator contributions, there is a
cancellation of the prefactor corresponding to the non-compact
4d momentum integration. This cancellation may not necessarily extend to
higher genus amplitudes (involving additional insertions
of closed string fields), although it certainly holds in the low energy
effective field theory limit~\cite{ov}.

The same amplitudes have been computed in the past, in the
context of Calabi-Yau compactifications of the heterotic string
and were shown to be identical to the partition function of  the
topological theory obtained by twisting the left-moving
supersymmetric sector which has $N=2$ superconformal symmetry~\cite{agnth}.
However, unlike in the Type II case, the recursion relations describing
the holomorphic anomaly of the heterotic topological partition
function do not close among themselves. They involve a new class
of topological quantities $F^{(0,h)}_n$, corresponding to correlation functions
of anti-chiral fields, that describe F-terms of the type
$\Pi^n (\Tr W^2)^{h-1}$, where $\Pi$'s are chiral projections of
non-holomorphic functions of $N=1$ chiral superfields.
Using Type I -- heterotic string duality~\cite{hetIdual}, one expects that the
heterotic topological partition function at genus $h-1$ coincides with
$F^{(0,h)}$ in the corresponding perturbative limit. On the Type I side, the
non closure of the recursion relations among $F^{(0,h)}$'s is due
to the existence of only one BRST charge on the boundaries, instead
of two acting separately on left and right movers.

An important property of $F^{(0,h)}$ with $h\ge 3$ is the breaking
of R-symmetry. A particularly interesting term is $(\Tr W^2)^2$.
Combined with supersymmetry breaking induced by vacuum
expectation values (VEV's) of R-preserving D-term auxiliaries,
it can generate Majorana gaugino masses,
$m_{1/2}\sim \langle D\rangle^2\sim m_0^4$ (in string units),
where $m_0$ is the scalar mass scale.
In this work, we present an explicit calculation of
$F^{(0,3)}$ on a world-sheet with three boundaries, in a simple
example of Type IIB compactified on $T^6$ with magnetized D9 branes,
or equivalently, upon T-duality, of intersecting D6 branes in Type IIA
orientifolds~\cite{Bach}. In a vacuum
configuration preserving $N=1$ supersymmetry, a non trivial
$F^{(0,3)}$ is generated if one of the three brane stacks associated
to the 3 boundaries is displaced away from the intersection of the
other two; this implies the breaking of R-symmetry in the above
example. On the other hand, if the brane configuration breaks
supersymmetry, Majorana gaugino masses are generated.
In the limit of small scalar masses, they behave as
$m_{1/2}\sim m_0^4$ with a coefficient given precisely
by the topological partition function $F^{(0,3)}$.
Similarly, the $\Pi{\rm Tr}W^2$ term associated to the topological
amplitude $F^{(0,2)}_1$ generates,
upon supersymmetry breaking, one-loop masses for non-chiral
brane fermions, of the same order as the gaugino masses.

Actually, the fact that a non-vanishing gaugino mass is generated
at the perturbative order $(g=0, h=3)$, which corresponds to Euler
characteristic $-1$, is in agreement with the general arguments
of Ref.\cite{at}, based on $N=2$ superconformal $U(1)$
charge conservation. It is the same order as $(g=1,h=1)$,
which corresponds
to the gravitational mediation of supersymmetry
breaking from the bulk to the gauge (brane) sector.
This contribution depends strongly on the mechanism
of supersymmetry breaking in the closed string sector~\cite{at},
ignored in the present work. From the effective field theory
point of view, the string diagram $(g=0, h=3)$ describes a
particular gauge mediation by
two-loop corrections in the gauge D-brane theory.\\[1ex]
The paper is organized as follows.

\noindent $\bullet$ In Section 2, we describe
the moduli space of the relevant world-sheet of genus zero with
$h$ boundaries, $(g=0,h)$, as obtained by an appropriate involution
of the Riemann surface of genus $g=h-1$ with no boundaries,
$(g=h-1,0)$~\cite{invo,modular}. We also discuss 
the partition function and the so-called correction factors
associated to Neumann and Dirichlet boundary conditions~\cite{modular}.

\noindent $\bullet$ In Section 3, we compute a string amplitude involving two
gauge fields and $2(h-2)$ gauginos on a world-sheet of
genus 0 with $h$ boundaries. We thus extract the coefficient
of the F-term $(\Tr W^2)^{h-1}$ in the effective action, as a function
of the closed string moduli parameterizing the Calabi-Yau
geometry. For simplicity, we perform the computation for
general $N=1$ intersecting brane configurations
in $N=2$ orbifold compactifications of Type II superstrings.
The generalization to Calabi-Yau is straightforward
using the results of Ref.\cite{agnt}.
We show that this coupling-coefficient is given by the
topological partition function $F^{(0,h)}$ of the associated twisted
Calabi-Yau sigma-model.

\noindent $\bullet$ In Section 4, we discuss the
holomorphic anomaly and the corresponding recursion relations.
We first review the situation in the heterotic theory
and then identify the possible contributions to the anomaly coming from
various degeneration limits (dividing geodesic and handle)
involving massless open and closed string states.

\noindent $\bullet$ In Section 5, we consider an explicit example
of Type I string compactifications on a factorized $T^6=(T^2)^3$
(instead of general orbifolds)  with
magnetized D9 branes, or equivalently with D6 branes at angles.
We then compute $F^{(0,3)}$ in the case where the angles are
chosen to preserve $N=1$ supersymmetry. We show
that a non-zero answer is obtained if one of the three brane
stacks associated to the three boundaries is moved away from
the intersection of the other two, in each of the three tori. For non-parallel
branes, this deformation does not change the spectrum,
but breaks all R-symmetries. $F^{(0,3)}$ is expressed in a compact
integral form as a function of $T^6$ moduli and Wilson lines.

\noindent $\bullet$ In Section 6, we compute the Majorana gaugino
masses for arbitrary brane angles that break supersymmetry,
originating from the same world-sheet with three
boundaries. We show that in the $N=1$ supersymmetric
limit, they behave as $m_0^4$, with $m_0$ the scalar
mass, with a coefficient given by the topological partition function
$F^{(0,3)}$, in accordance with the result of Section 5.

\noindent $\bullet$ In Section 7, we compute certain matter mass terms
that are related to the F-terms appearing in the holomorphic
anomaly of the gaugino mass $F^{(0,3)}$. These terms involve (twisted)
charged fields from brane intersections.

\noindent $\bullet$ Section 8 contains some explicit examples.
In particular, we consider the case of three stacks of branes forming
pairwise three different $N=2$ supersymmetric sectors.
We can then evaluate the $F^{(0,3)}$ integral  explicitly, as well as
all quantities related to it by the holomorphic anomaly, namely
$F^{(0,2)}_1$ and $F^{(0,1)}_2$.

\noindent $\bullet$ Our results are summarized in Section 9 which also
contains discussions on possible applications and open problems.

\noindent $\bullet$ Finally in the Appendix, we derive the lattice contribution
of the twisted bosons to the topological partition functions $F^{(0,h)}$
for the case of magnetized D9 branes. This involves periods associated with
Prym differentials and to our knowledge a detailed treatment of the zero modes
associated with the twisted bosons for the open string case has not been made
in the physics literature (for the closed string case see Ref.\cite{nsv}). The
result of this Appendix for the special case $h=3$ is used in sections 5 and 6.

\section{Genus $g=0$ World-Sheet with $h$ Boundaries From the Involution of
$g=h-1$ Riemann Surface}\label{hsurface}

A planar ($h{-}1$)-loop open string world-sheet with $h$ boundaries,
see Fig.\ 1,
\begin{figure}[h]\hfill
\includegraphics[scale=.7]{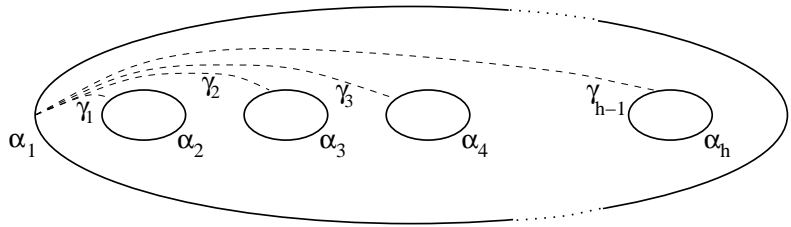}\hfill\hfill
\caption{Open string world-sheet with $h$ boundaries.}
\label{topf1}
\end{figure}
admits a
closed orientable genus $h{-}1$ double cover. The embedding is described by an
anti-conformal involution, with the boundaries made of its fixed points. In the canonical
homology basis, the {\bf a}-cycles are invariant, ${\bf a}_i\to{\bf \bar a}_i={\bf a}_i$,
while the {\bf b}-cycles have their orientation reverted, ${\bf b}_i\to{\bf\bar  b}_i={\bf
b}_i^{-1}$, see Fig.\ 2.
\begin{figure}[h]\hfill
\includegraphics[scale=.6]{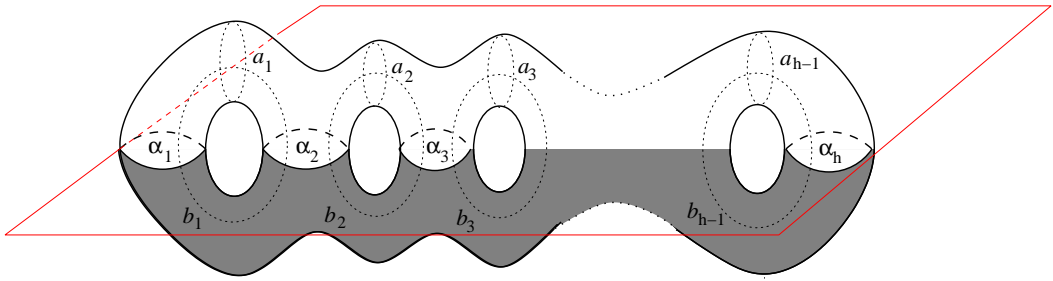}\hfill\hfill
\caption{
The involution (\ref{invh}) acts on the $g=h{-}1$ double-cover as a mirror
reflection in the horizontal plane. The open string world-sheet is in the
upper half while its mirror image is in the shaded lower half.}
\label{topf2}
\end{figure}
The anti-symplectic involution matrix has the form
\be
I=\left(\begin{matrix}\id&0\cr 0& -\id\end{matrix}\right),
\label{invh}
\ee
where $\id$ is the $(h-1)\times (h-1)$ identity matrix.
In terms of the
{\bf a}-cycles of the double cover, the boundaries of the $(g=0,h)$ world-sheet are
${\bf\alpha}_1={\bf a}_1,\dots,\alpha_k={\bf a}_k\,{\bf a}_{k-1}^{-1},\dots,\alpha_h={\bf
a}_{h-1}^{-1}$. It can be made into a contractible region by cutting along the curves
$\gamma_i$, $i=1,\dots,h-1$, as shown in Fig.\ 1. The corresponding fundamental polygon is
shown in Fig.\ 3.
\begin{figure}[h]\hfill
\includegraphics[scale=.3]{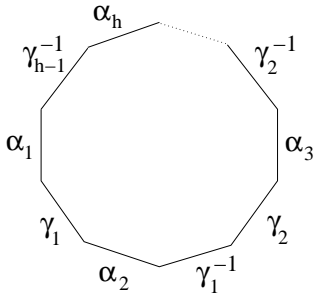}\hfill\hfill
\caption{The fundamental polygon obtained by cutting the open string
world-sheet along the curves shown in Figure 1.}
\label{topf3}
\end{figure}
Under the involution, $\gamma_i\to\bar\gamma_i$, such that
$\gamma_i\,\bar\gamma_i^{-1}={\scriptstyle \prod}_{k=1}^{k=i}{\bf b}_k$.

The period matrix $\tau$ of the double cover is restricted by the condition of
invariance under the involution: ${\bar\tau}=I(\tau)=-\tau$, thus
\be
\tau=it\, ,
\label{periodm}
\ee
where $t$ is a
real, symmetric matrix. Note that the positivity requirement for the period matrix now reads,
in particular, $\det t>0$. The modular transformations that preserve the involution are
represented by $Sp(2g, \mathbb{Z})$ matrices of the form
\be
M=\left(\begin{matrix}(D^{-1})^T&0\cr 0& D\end{matrix}
\right).
\ee
Since $D$ and $D^{-1}$ must be integer-valued, the
``relative modular group" \cite{modular} is $GL(g,\mathbb{Z})$. This group essentially
interchanges the boundaries, transforming the period matrix as $\tau\to D\tau D^T$.
Note that since $\det D=\pm 1$, the determinant $\det t$ remains invariant.

The double cover endows the world-sheet with the basis of holomorphic differentials
$\omega_i$ normalized as
\be
\int_{{\bf a}_j}\omega_i=\delta_{ij} \quad ,\quad \int_{{\bf
b}_j}\omega_i=it_{ij},
\label{norms}
\ee
The involution transforms them into
anti-holomorphic differentials
\be
\overline{\omega_i(z)}=\bar\omega_i(\bar z)\quad ;\quad
\int_{\gamma_j}\omega_i-\bar \omega_i=i\sum_{k=1}^{k=j}t_{ik}.
\ee
Furthermore, the
normalization condition (\ref{norms}) implies
\be
\int_{{\alpha}_m}\omega_n=\delta_{mn}-
\delta_{m(n-1)}\, .
\label{ibound}
\ee
For future use, we also need the surface integrals
\be
\int\omega_i\wedge{\bar\omega}_j=\int
d\left(\int^z\!\!\omega_i\right) ({\bar\omega}_j-\omega_j)
\ee
which, after using the fundamental polygon with the boundary conditions
${\bar\omega}_j|_{\alpha_k}=\omega_j|_{\alpha_k}$, become
\be
\int\omega_i\wedge{\bar\omega}_j=\frac{i}{2}\sum_{k=1}^{k=h-1}\int_{\alpha_{k+1}}
\!\!\!\omega_i\int_{\gamma_{k}} ({\bar\omega}_j-\omega_j
)=\frac{t_{ij}}{2}.
\label{intij}
\ee
In particular, the above result implies
${t_{ii}}>0$ for the diagonal elements of the period matrix.

The coordinates of a string propagating on a circle of radius $R$ have the form
\be
X(P)=2\pi R\sum_{i=1}^{i=g}[L_i\int^P_{P_0}\omega_i(z) +
\bar L_i\int^P_{P_0}\bar\omega_i(\bar z)]\,
,\ee
where $P_0$ is an arbitrary base point. Due to the reality property
${\bar\omega}_j|_{\alpha_k}=\omega_j|_{\alpha_k}$,
the Neumann (N) and Dirichlet (D) boundary
conditions read, respectively,
\be
{\bar L}_{i}=\pm L_{i}\qquad \left\{ {+~{\rm for}~N\atop
-~{\rm for}~D}\right.
\ee
A string satisfying Neumann boundary conditions can wind around the
boun\-daries $\alpha_i$. Keeping in mind the double cover, it is convenient to parameterize
these windings in terms of the winding numbers $n_i$ around ${\bf a}_i$:
\be
L_i^N=\frac{n_i}{2}\, .
\ee
On the other hand, in the Dirichlet directions, the boundary
strings cannot wind, but an open string stretching between two different boundaries can wind
from one end to the other.  If it winds $m_i$ times between $\alpha_i$ and $\alpha_{i+1}$,
then
\be
L_i^D=i\sum_{j}(t^{-1})_{ij}m_j\, .
\ee
In terms of the double cover, $2m_i$
corresponds to the (even) number of windings around ${\bf b}_i$. The classical action,
$S_{\rm cl}={1\over \pi\alpha'}\int\partial_zX\partial_{\bar z}X$, can be
computed by using (\ref{intij}):
\be
S_{\rm cl}^N=\frac{\pi R^2}{2}\,n t \,
n^T\qquad,\qquad  S_{\rm cl}^D = 2\pi R^2\, m t^{-1} m^T\, ,
\label{actions}
\ee
where $R$ is the compactification radius in $\alpha'$ units.

The full partition function, in addition to the lattice sum $Z_{\rm cl}=e^{-S_{\rm cl}}$,
includes the bosonic determinant factors. These can be obtained by taking the appropriate
square root of the corresponding expression for the double cover:
\be
(\det{\rm Im}\tau)^{-1/2}|Z_1|^{-1}\quad \longrightarrow\quad
(R_{\Sigma}\det{\rm Im}\tau)^{-1/4}Z_1^{-1/2}\, ,
\label{dubdet}
\ee where $Z_1$ is the chiral part of the
determinant and $R_{\Sigma}$ is the ``correction factor'' depending on the boundary
conditions~\cite{modular}. For a general involution, it has the form
\be
I=\left(\begin{matrix}\scriptstyle
\Gamma&\scriptstyle  0\cr \scriptstyle \Delta&\scriptstyle \Gamma\end{matrix}\right)\,
:\quad R_{\Sigma}^{\rm N}=(R_{\Sigma}^{\rm D})^{-1}= \det\left({1-\Gamma\over 2}{\rm
Im}\tau+ {1+\Gamma\over 2}({\rm Im}\tau)^{-1}\right). \label{corgen}
\ee
Thus in our case,
\be
R_{\Sigma}^{\rm N}=(\det t)^{-1}\qquad{\rm and}
\nonumber
\ee
\be
(\det{\rm Im}\tau)^{-1/2}|Z_1|^{-1}\quad \longrightarrow\quad \left\{
{Z_1^{-1/2}~\;\qquad\qquad ~{\rm for}~N\atop (\det t)^{-1/2}Z_1^{-1/2} ~{\rm
for}~D}\right.
\label{corfin}
\ee

\section{$W^{2(h-1)}$ F-terms from Open String Topological Amplitudes}

In Type II theory, the F-terms of the form ${\cal W}^{2g}$, where  $\cal W$ is the  chiral
$N=2$ supergravity multiplet, are determined at genus $g$ by the topological partition
function $F_g$~\cite{bcov,agnt}. In this
section, we discuss a similar structure in Type I theory: the open string diagrams with $h$
boundaries ($h{-}1$ open string loops) generate the effective action terms
$(\Tr\epsilon^{\alpha\beta}W_{\alpha}W_{\beta})^{(h-1)}$, where $W$ is the familiar chiral
(spinorial) gauge field strength superfield and the trace is  over  gauge indices in the
fundamental representation \cite{bcov,ov,bov}. We will demonstrate this fact by evaluating the amplitudes
involving $2(h{-}2)$ gauginos and two gauge bosons coupled to $h$ boundaries of a genus zero
surface, in the Neveu-Schwarz-Ramond formalism, similarly to Ref.\cite{agnt}. It is very convenient to describe this surface in terms of its $g=h{-}1$ double
cover introduced in the previous section. Then the computation proceeds by essentially
repeating the steps of the original computation in Type II theory.

We are interested in the effective action term $(\left.
\Tr\epsilon_{\alpha\beta}W_{\alpha}W_{\beta})^{(h-1)}\right|_{\rm
F}=(h{-}1)\Tr[(\epsilon^{\alpha\beta}\lambda_{\alpha}
\lambda_{\beta})^{(h-2)}g^{ac}g^{bd}{\cal F}_{ab}{\cal F}_{cd}]+\dots$, where $\lambda$ are
gauginos and $\cal F$ are the self-dual combinations of gauge field strengths. Thus the
amplitude under consideration involves $h{-}2$ pairs of gauginos at zero momentum, with
opposite helicities inside each pair, plus one pair of gauge bosons in the momentum-helicity
configuration corresponding to a self-dual gauge field strength. In order to provide the
desired gauge charge, helicity and momentum configuration, the gaugino vertex operators are
distributed in pairs among $h{-}2$ boundaries, the two gauge boson vertices are inserted on
a separate boundary while one boundary remains ``empty''.

The gaugino vertex operator of definite helicity $\alpha$, at zero momentum, in the
canonical $-1/2$ ghost picture, reads:
\be V^{(-1/2)}_\alpha(x)=:e^{-\varphi/2}S_\alpha
S_{int}: \, ,
\label{lvertex}
\ee
where $x$ is a position on the boundary of the world-sheet,
$\varphi$ is the scalar bosonizing the superghost system, and $S_\alpha$ ($S_{int}$) is the
space-time (internal) spin field. Upon complexification of the four space-time fermionic
coordinates, $\psi_I$ for $I=1,2$, and introducing the bosonization scalars
$e^{i\phi_I}=\psi_I$, one has
\be
S_\alpha =e^{\pm\displaystyle {i\over 2}(\phi_1 +\phi_2)}
\quad ;\quad\alpha=\pm \, .
\label{spinf}
\ee
The gauge boson vertex operator at momentum $p$
and polarization $\epsilon$, in the $0$ ghost picture, is:
\begin{equation}
V_{p,\epsilon}^{(0)}(x) ~=~ :\!\epsilon_{\mu} (\partial X^{\mu}+ip\cdot\psi
 \psi^{\mu})
e^{ip\cdot \!X}\!:~.
\label{vg}
\end{equation}
The above vertices must be supplemented by the appropriate Chan-Paton factors,
which we omit here for simplicity.

In order to balance the ghost charge, we change one half of gaugino vertex operators to
$+1/2$ ghost picture: this is done by inserting $h{-}2$ picture changing operators at the
boundaries. Recall that the picture changing operator (PCO) is defined as $e^{\varphi}T_F$,
where $T_F$ is the world-sheet supercurrent. In addition, due to the supermoduli
integration, there are the usual $2g{-}2$ PCO insertions on the genus $g$ double cover that
can be realized as $2(h{-}2)$ boundary insertions on the open string world-sheet. As a
result, the total number of PCO insertions is $3(h{-}2)$.

We will see below that the purely bosonic parts of the gauge boson vertex operators do not
contribute to the amplitude under consideration. To make the calculation simpler, we choose
a kinematical configuration corresponding to $\psi_1\psi_2$ from one vertex and to
$\bar\psi_1\bar\psi_2$ from the second. The amplitude then becomes
\be
A_h=\langle\prod_{i=1}^{h-2} e^{-\varphi/2} S_+S_{int}(x_i)\prod_{i=1}^{h-2} e^{-\varphi/2}
S_-S_{int}(y_i)\, \psi_1\psi_2(z) {\bar\psi}_1{\bar\psi}_2(w) \!\! \prod_{i=1}^{3(h-2)}\!\!
e^{\varphi} T_{F}(z_i){\rangle}_h
\label{ampl}
\ee
The above expression has exactly the
same structure as the left-moving (or right-moving) part of the topological amplitude
written in Eq.(3.6) of Ref.\cite{agnt}. The amplitudes of Ref.\cite{agnt} describe
scattering processes involving graviphotons and gravitons; here, these particles are
replaced by gauginos and gauge bosons, respectively. Now the vertex positions $x_i$, $y_i$,
$z$ and $w$ are integrated over the boundary while the supercurrents are inserted at {\em a
priori} arbitrary points $z_i$ of the boundary. In order to demonstrate a similar,
topological nature of $A_h$, we can limit our considerations to the case of orbifold
compactifications.
The twists around {\bf b}-cycles of the double cover correspond to the brane angles, while the twists
around {\bf a}-cycles belong to the orbifold group of the Type II theory. Thus, the former
are fixed by the brane configuration, while the latter are summed over all elements
of the orbifold group. Below, we call both types of periodicity conditions as orbifold twists.

In the case of orbifolds, the internal $N=2$ SCFT is realized in terms of free bosons and
fermions. We consider for simplicity orbifolds realized in terms of $3$ complex bosons
$X_{I}$ and left- (right-) moving fermions $\psi_{I}$ ($\widetilde{\psi}_{I}$), with
$I=3,4,5$. Since all vertices involving these fermions are inserted at the boundary, from
now on we can identify the left- and right-movers. Let $h_O$ be an orbifold
twist defined by $h_O=\{h_{I}\}$, and its action on $X_I$ is $X_{I}\rightarrow e^{2\pi i
h_{I}}X_I$ and similarly for $\psi_I$.  Space-time supersymmetry implies that one can always
choose the $h_I$'s to satisfy the condition:
\begin{equation}
\sum_{I} h_{I}~=~0\ .
\label{susy}
\end{equation}
On the genus $g=h{-}1$ double cover, we must associate one orbifold twist
to each homology cycle ${\bf a}_{i},{\bf b}_{i}$, for $i=1,\cdots,h{-}1$. In the
following we shall denote by $\{h_O\}=\{\{h_I\}\}$ the set of all twists along different
cycles. One can bosonize the complex fermions
\begin{equation}
\psi_{I}=e^{i \phi_I}~,~~~~~~~~~~~~~\bar{\psi}_{I}=e^{-i\phi_I}~.
\label{boso}
\end{equation}
In terms of these bosons, the internal part of gaugino vertex operators reads
\be
\label{spinfint}
S_{int} =e^{\displaystyle {i\over 2}(\phi_3 +\phi_4 +\phi_5)} \, .
\ee
Similarly, the internal part of the supercurrent at the boundary becomes
\be
T_F^{int}= G^- + G^+ \quad; \quad  G^-= \sum_{I=3}^5  e^{-i\phi_I}\partial X^I\, .
\label{tfint}
\ee
By internal charge conservation, since all $2(h{-}2)$ gauginos carry charge $+1/2$ for
$\phi_3$, $\phi_4$ and $\phi_5$, only the internal parts (\ref{tfint}) of the supercurrents
contribute: $h{-}2$ $T_F$'s must contribute $-1$ charge each for $\phi_3$ and similarly $-1$
for $\phi_4$ and $\phi_5$ each. Thus only $G^-$ contributes in this amplitude.

In order to compute the amplitude \ref{ampl}, we repeat the steps leading from Eq.(3.6) to
Eq.(3.18) of Ref.\cite{agnt}. In particular, in order to cast the contribution of
world-sheet fermions into a form that makes the summation over their spin structures
tractable by using the simplest form of Riemann identity for theta functions, we are led to
the following choice of the supercurrent insertion points:
 \begin{equation}
\sum_{a=1}^{3(h-2)}z_{a}~=~\sum_{i=1}^{h-2}y_{i}-z+w+2\Delta~,
\label{gauge}
\end{equation}
where $\Delta$ is the Riemann $\theta$ constant associated to the double cover. Note that
this is an allowed gauge choice even if all points are located at the boundary. After summing
over all spin structures and using exactly the same bosonization formulae and theta function
identities as in section 3 of Ref.\cite{agnt}, we obtain the following expression
\begin{eqnarray}
A_{h}~&=&\det{\omega_{i}(x_j,z)} \det{\omega_{i}(y_{j},w)} \frac{\prod_{I}\det{(\partial
X_{I}\omega_{-h_I,i} (u_{j,I}))}}{\det h_{a}(z_{b})}\nonumber\\ & &
\epsilon_{a_{1},\cdots,a_{3(h-2)}}\prod_{i=1}^{3(h-2)} \int (\mu_{i}h_{a_i})
\,({\makebox{lattice sum}})~,
\label{top}
\end{eqnarray}
where $\partial X_{I}$ are the instanton modes twisted along the homology cycles by a
particular set of orbifold twists $\{h_I\}$ and $\omega_{-h_I,i}$, $i=1,\dots,h-2$,
are the differentials associated to the boundary conditions twisted by $\{e^{-2i\pi h_I}\}$.
The above expression is written for one particular partition  $u_{i,I}~ (i=1,\dots,h-2)$,
$I=3,4,5$ of the positions of $T_F$'s which contribute $-1$ charge for $\phi_3$, $\phi_4$
and $\phi_3$, respectively. At the end, one must consider all possible partitions
$\{u_{i,I}\}$ and antisymmetrize; note that as a set $\{z_a\}=\bigcup_{i,I} \{u_{i,I}\}$.
Finally, $h_a$, $a=1,\dots ,3(h-2)$ are the $3(h-2)$ quadratic differentials whose
determinant appears after a number of technical steps explained in Ref.\cite{agnt}, making
use of the bosonization formulae. They also appear as the zero modes of $b$-ghosts,
contracted with the Beltrami differentials $\mu_i$; the corresponding integral gives the
measure over moduli space of genus zero surfaces with $h$ disconnected boundaries.

There is only one difference between Eq.(\ref{top}) and the analogous Eq. (3.18) of
Ref.\cite{agnt}: the absence of $(\det{\rm Im}\tau)^{-2}$ factor which arises in Type II
theory after integrating over the space-time zero modes $X^{\mu}$ ({\em i.e.} the
four-dimensional momenta). In our case, the correction factor (\ref{corgen}) for the Neumann
boundary condition eliminates this determinant, {\em c.f.} Eq.(\ref{corfin}). Furthermore,
for each compact direction, there is a lattice weight $Z_{cl}=e^{-S_{cl}}$, see
Eq.(\ref{actions}), which according to Eq.(\ref{corfin}) should be multiplied by
$(\det t)^{-1/2}$ only in the case of Dirichlet boundary conditions.

The result (\ref{top}) is for a fixed partition $\{u_{i,I}\}$ of the $z_a$'s. As mentioned
earlier one must consider all possible partitions and antisymmetrize. Furthermore, $\partial
X_{I}\omega_{-h_{I},i}$ are holomorphic quadratic differentials. Therefore, summing aver all
partitions $\{u_{i,I}\}$ with the proper antisymmetrization gives:
\begin{equation}
\sum_{\{u\}}\prod_{I} \det(\partial X_{I}\omega_{-h_{I},i} (u_{i,I}))~=~B \det
h_{a}(z_b)~,
\label{det}
\end{equation}
where $B$ is $z_a$-independent. In this way, the amplitude (\ref{top}) becomes manifestly
independent of the PCO's insertion points.

Now it remains to integrate the vertex positions $x_i$, $y_i$, $z$ and $w$ over all $h$
disconnected boundaries $\alpha$. For $z$ and $w$ located at a specific boundary $\alpha_k$,
each pairing $(x,y)$ of opposite helicity gauginos at other boundaries gives rise to a
specific Chan-Paton factor and, as a consequence of Eq.(\ref{ibound}), it picks up only one
of the $(h-1)!^2$ terms from the product $\det{\omega_{i}(x_j,z)}
\det{\omega_{i}(y_{j},w)}$. The integral of such a term is simply 1. As a result,
\begin{equation}
A_{h}~=~N\,h!\int_{{\cal M}_h}B \det{\int({\mu_a}h_{b}}) \,({\rm
{lattice~sum}})~,
\label{top1}
\end{equation}
where ${{\cal M}_h}$ is the moduli space of genus 0 Riemann surfaces with $h$
boundaries.\footnote{Strictly speaking one descends on ${{\cal M}_h}$ after averaging over
periodicity conditions of the orbifold group around the {\bf a}-cycles.}
The combinatorial factor $h!$ arises as follows. First, there are $h$
choices of the ``empty'' boundary, then $h{-}1$ choices of the gauge bosons' boundary and
finally, $(h{-}2)!$ ways of distributing say positive helicity gauginos -- the positions of
negative helicity gauginos determines the Chan-Paton factor. The additional factor $N$ comes
from the ``empty'' boundary and counts the number of D-branes.

The effective action term that reproduces the amplitude (\ref{top1}) is the F-term
\be
S_{\rm eff}=\left.
F^{(0,h)}(\Tr\epsilon^{\alpha\beta}W_{\alpha}W_{\beta})^{(h-1)}\right|_F,
\ee
with the coefficient given by the topological partition function
\begin{equation}
F^{(0,h)}~=~N\,h\int_{{\cal M}_h}\prod_{a=1}^{3h-6}{\int({\mu_a} G^-}) \,({\rm
{lattice~sum}})\, ,
\label{ftoph}
\end{equation}
where we have used Eqs.(\ref{det}) and (\ref{top1}).  Here $G^-= \sum_I \partial X^I
\bar{\psi}^I$ is the supercurrent in the topological twisted theory. Note that in the
latter $\bar{\psi}^I$ carry dimension 1
and therefore they have $h-2$ zero modes $\omega_{-h_I,i}$ with $i=1,...,h-2$.

\section{Holomorphic Anomaly and $\Pi$-terms}

Type I models considered above are dual to heterotic models. In particular
the $N=1$ Type I models should be dual to heterotic models based on an
$N=(2,0)$ world-sheet superconformal field theory. Correspondingly the
topological amplitudes giving rise to F-terms $F^{(0,h)}({\rm Tr}W^2)^{h-1}$ in
Type I has a counterpart in heterotic theory. In Ref.\cite{agnth}, such amplitudes
were shown to be given by partition functions of topological heterotic
theory obtained by twisting the left-moving $N=2$ superconformal algebra.
Indeed genus $g$ partition function of the topological heterotic theory $F^H_g$
gives the coupling $F^H_g ({\rm Tr}W^2)^g$.

The topological partition functions in Type II theories satisfy a holomorphic
anomaly equation; their derivatives with respect to anti-holomor\-phic moduli
can be expressed in terms of topological partition functions of lower genera.
In the heterotic theory, however, as shown in Ref.\cite{agnth}, anti-holomorphic
derivatives of $F^H_g$ give rise to a larger class of ``topological" quantities
$F^H_{g,n}$ labeled by genus $g$ and insertions of $n$ pairs of anti-chiral
operators. In the effective field theory, these quantities are associated to
F-terms of the form $({\rm Tr}W^2)^g \Pi^n$. Let us briefly
recall this difference between the Type II and heterotic topological theories.

The left-moving $N=2$ superconformal algebra is generated by the stress tensor
$T$, a $U(1)$ current $J$ and the two dimension $3/2$ superconformal
generators $G^{+}$ and $G^{-}$ where the superscripts $\pm$ refer to their
$U(1)$ charges. Topological twisting of this algebra amounts to modifying
$T \rightarrow T + \frac{1}{2}\partial J$. With respect to the new $T$ the
dimensions of $G^+$ and $G^-$ are respectively $1$ and $2$. One defines the
BRST operator for the topological theory $Q_{BRST}$ as the contour integral
of $G^+$ current. $G^-$ having dimension $2$ and satisfying the condition
$Q_{BRST} G^- = T$ plays the role of the $b$ ghost
field of the string theory. Thus the measure on the moduli space of a
Riemann surface of genus $g$ is defined by computing the correlation function of $3g-3$
$G^-$'s which are each folded with a Beltrami differential. The physical states of
the theory are given by the $Q_{BRST}$-cohomology; the chiral primaries $\phi_i^+$
having charge $+1$ have now zero dimension, they are in $Q_{BRST}$-cohomology and
can be inserted at punctures on the Riemann surfaces, while the anti-chiral
vertex operators (in the 0 picture in string theory) are BRST exact
operators $Q_{BRST} \phi_{\bar{i}}^- = \oint G^+ \phi_{\bar{i}}^-$. Inserting such an
operator in the Riemann surface gives a total holomorphic derivative in the world-sheet moduli
$\partial_t$, since by deforming the contour $Q_{BRST}$ will act on
one of the $G^-$ folded
with the Beltrami differentials and convert it to $T$.  Thus the only possible
contributions can come from the boundaries of the moduli space of the world-sheet,
namely the degenerations of the original Riemann surface along some non-trivial
cycle which could be either homologically trivial or non-trivial.

In the homologically trivial degeneration limit,
the surface splits into two Riemann surfaces $\Sigma_1$ and $\Sigma_2$ of lower genera,
say genus $g_1$ and $g_2$, with one puncture each at $P_1$ and $P_2$.
The operator $\phi_{\bar{i}}^-$ will be on one of the surfaces (say $\Sigma_1$). Since in the
twisted theory the total charge on the sphere must be +3, it follows that the operators that
appear at $P_1$ and $P_2$ are dual of each other and carry charges +2 and +1, respectively.
The Beltrami differentials together with $G^-$ split on the two surfaces according to
the $U(1)$ charge conditions in the twisted theory. The resulting term is of the form
\begin{equation}
F^{g_1}_{\bar{i},\bar{j}}G^{\bar{j} j} D_j F^{g_2}\, ,
\label{trivdeg}
\end{equation}
where  $D$ is the holomorphic covariant derivative, and $j$ denotes the chiral operator
carrying charge +1 at $P_2$, while the operator at
$P_1$ carrying charge +2 is related to the anti-chiral operator (with charge $-1$) labeled
by $\bar{j}$ by the action of a holomorphic 3-form operator $\rho(z)$ carrying charge +3:
\be
\phi_{\bar{j}}^-\to\phi_{\bar j}^{++}\equiv\oint dz\rho(z)\phi_{\bar{j}}^-\, .
\label{rho}
\ee
Such an operator $\rho$ of dimension 0 in the twisted theory exists for all $N=2$
superconformal field theories leading to space-time supersymmetry: $\rho=e^{i{\sqrt 3}H}$
with ${\sqrt 3}\partial H$ being the $U(1)$ current. The metric $G_{j\bar j}$ appearing in
Eq.(\ref{trivdeg}) is defined by the inner product $\langle\phi_j^+\phi_{\bar j}^{++}\rangle$.
In Ref.\cite{agnth}, these new topological objects $F^g_{\bar{i},\bar{j}}$ were related to
the couplings of the effective action terms $({\rm Tr}W^2)^g\Pi$,
denoted symbolically $F^g_1$.
By further taking anti-holomorphic derivatives
of these terms one arrives at generalized topological quantities denoted by
$F^g_{\bar{i}_1...\bar{i}_n ; \bar{j}_1...\bar{j}_n}$ where $\bar{i}_k$ denote charge $-1$
insertions while $\bar{j}_k$ denote charge +2 insertions. These quantities were identified
with the effective action terms $({\rm Tr}W^2)^g (\Pi)^n$.

Similar reasoning for the case of handle degeneration results in a term of the form
\begin{equation}
C^{j \bar{j}} D_j F^{g-1}_{\bar{i},\bar{j}}
\end{equation}
with
\begin{equation}
C^{j \bar{j}} = G^{j\bar{k}} Q_{\bar{k}\ell} G^{\ell \bar{\ell}} Q_{\bar{\ell} k}
G^{k \bar{j}}\, ,
\label{Cs}
\end{equation}
where $Q$ is the charge operator associated with the gauge field $W$ (recall that in the
heterotic theory the non-topological right moving part of the gauge vertex contains the
charge operator). This is because only
$2g-2$ of the original $2g$ gauge superfields sit on the genus $g-1$ surface obtained from
the handle degeneration and the remaining two $W$'s sit at the node. As a result, the
propagator connecting the two punctures $P_1$ and $P_2$ comes with the charges of the two
$W$'s.\footnote{More generally there
can be several different gauge fields and the corresponding topological partition function will
carry the labels of the gauge fields. In this case the $Q$'s in equation (\ref{Cs}) will
carry the labels of the two gauge fields that sit at the node.}

In Type II, we have also the right-moving $N=2$ superconformal
algebra, which upon twisting gives rise to another BRST operator $\bar{Q}_{BRST}$.
The anti-chiral vertex operators now (in the (0,0) ghost picture) are of the form
$Q_{BRST} \bar{Q}_{BRST} \phi^{--}$ and inserting it gives rise to a double derivative
$\partial_t \partial_{\bar{t}}$ in the moduli space integrals, where $t$ is standard
plumbing fixture coordinate describing the degeneration of the surface. To get a non-
vanishing contribution at $t \rightarrow 0$ the integrand must behave as $\ln(t\bar{t})$.
The latter behavior is obtained only if the operator $\phi^-$ approaches the node
(the puncture $P_1$) and has a non-vanishing structure constant $C_{\bar{i}\bar{j}\bar{k}}$
for some $\bar{k}$,
the integral of its position giving rise to $\ln (t\bar{t})$. As a result
\begin{equation}
F^{g}_{\bar{i},\bar{j}} = e^{2K}C_{\bar{i}\bar{j}\bar{k}} G^{\bar{k}k}D_k F^g\, ,
\end{equation}
where $K$ is the K\"ahler potential
(the appearance of $e^{2K}$ can be deduced by matching the K\"ahler weights). Moreover for
the handle degeneration case $C^{j \bar{j}}$ in Eq.(\ref{Cs}) reduces to the inverse metric
$G^{j \bar{j}}$. This is because
the charge operator $Q$ is essentially replaced by space-time momenta which are part
of the graviphoton field strength in the  $N=2$ $({\cal W}^2)^g$ term.

Returning to the case of open strings (Type I) we expect the situation to be similar to the
heterotic string since firstly we are dealing with $N=1$ theories and secondly Type I and
heterotic
theories are dual to each other. This can be seen more directly by examining the corresponding
topological theories. Indeed on surfaces with boundaries, only the sum of the left
and right moving BRST operators is compatible with the boundary conditions. As a result, upon
inserting an anti-chiral operator in the topological partition function one can only deform
the contour associated with this combination of left and right BRST operators resulting in
an integrand which is a total derivative
in the world-sheet moduli space. There is no second BRST operator which could give an
additional total derivative. To study the possible boundaries of the moduli space
(degenerations) let us, for simplicity, restrict ourselves to surfaces of genus 0 with $h$
boundaries $\Sigma_{(0,h)}$. In the following we assume $h \geq 3$. The moduli space of such
surface is $3(h-2)$ real dimensional implying that there are $3(h-2)$ insertions of the sum of
left and right moving $G^-$ that are folded with the Beltrami differentials. This surface
therefore carries a (left plus right) $U(1)$ charge equal to $3(h-2)$ which is exactly the
number dictated by the $U(1)$ anomaly in the twisted theory. Inserting an anti-chiral
operator and deforming the contour associated with the corresponding BRST operator converts
one of the $G^-$ into the stress-tensor, giving rise to a total derivative in the
moduli space. The boundary terms come from the degeneration limits of the surface. There are
two degenerations with open string intermediate states analogous to the dividing and handle
degeneration cases of heterotic string, and one with intermediate closed string (see Fig.~4).
 \begin{figure}[h]\hfill
 \includegraphics[scale=.34]{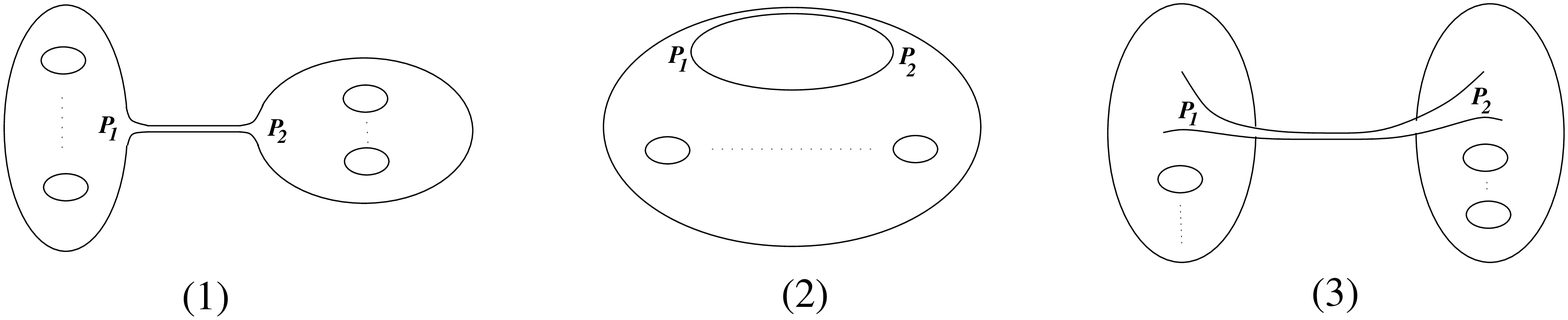}\hfill\hfill
 \caption{The three degeneration limits.}
 \label{topf5}
 \end{figure}

\begin{enumerate}
\item The first degeneration arises when the surface splits into two Riemann surfaces
$\Sigma_{(0,h_1)}$ and $\Sigma_{(0,h_2)}$ (where $h=h_1+h_2-1$ and $h_1, h_2 \geq 2$)
with one puncture each, say $P_1$ and $P_2$, at the
boundaries of the two surfaces. 
The original anti-chiral insertion $\Phi_{\bar{i}}^-$
is on one of the components, say $\Sigma_{(0,h_1)}$. Since the total charge (i.e. left plus right
$U(1)$ charge) on a disk in the twisted theory, as dictated by the $U(1)$ anomaly,
should be +3, it follows that the open string insertion at $P_1$ carries
charge +2, while the open string insertion at $P_2$ is its metric-dual operator $\phi_j^+$
carrying charge +1. The charge +2 state is obtained now by the action of $\rho_L+\rho_R$ on
a charge $-1$ operator. The
leftover $3(h-2)-1$ (left plus right) $G^-$'s folded with the Beltrami differentials distribute
themselves on $\Sigma_{(0,h_1)}$ and $\Sigma_{(0,h_2)}$. Their numbers on the two surfaces
are respectively $3(h_1-2)+1$ and $3(h_2-2)+1$ which is dictated by the $U(1)$ charge
anomalies on the two surfaces, as well as by the dimension of the moduli space of the
corresponding one punctured Riemann surfaces. The resulting term is
\begin{equation}
F^{(0,h_1)}_{\bar{i},\bar{j}}G^{\bar{j} j} D_j F^{(0,h_2)}  ; ~~~~~h_1+h_2 = h+1\, .
\label{deg1}
\end{equation}

\item The second open string degeneration is analogous to the handle degeneration of the
closed string and results in twice-punctured surface $\Sigma_{(0,h-1)}$ with punctures $P_1$
and $P_2$ on a boundary with the intermediate states at the two punctures carrying charge +2
and +1 respectively. The dimension of this twice-punctured moduli space is $3(h-3)+2$ which
is exactly the number of the leftover  $G^-$'s. This is also in agreement with the total
$U(1)$ charge anomaly. The resulting term is
\begin{equation}
C^{j \bar{j}} D_j F^{(0,h-1)}_{\bar{i},\bar{j}}\, .
\label{deg2}
\end{equation}

\item Finally there is also a degeneration with closed string intermediate state when the
surface splits into $\Sigma_{(0,h_1)}$ and $\Sigma_{(0,h_2)}$ with one puncture each $P_1$
and $P_2$ in the interior of the two surfaces. It is easy to see that $h_1+h_2=h$ (where
now $h_1, h_2\ge 1$). The plumbing fixture coordinate $\tau$ is now complex and the boundary
corresponds to $t=|\tau|\to\infty$. Thus the angular part of $\tau$ is still to be integrated.
The moduli space of the two surfaces
with one puncture each in the interior of the surfaces is $3(h_1-2)+2$ and $3(h_2-2)+2$ real
dimensional, respectively, implying that as many $G^-$'s are distributed on the two surfaces,
respectively. The remaining one $G^-$ is sitting at the node which is folded with the
Beltrami differential corresponding
to the angular part of $\tau$. Now the question is what are the $U(1)$ charges
carried by the closed string intermediate state at $P_1$ and $P_2$. Since the $U(1)$ anomaly
for a sphere (which is the relevant surface for the intermediate closed string propagation)
is +6, we conclude that the sum of the $U(1)$ charges at $P_1$ and $P_2$ is +6.
If $\phi_{\bar{i}}^-$ is on the first surface, then there are two possibilities: the charges at
$(P_1, P_2)$ are $(+4,+2)$ or $(+3,+3)$. The charges here are the sum of the left and right
moving parts. Charge +4 therefore means (left, right) (anti-chiral, anti-chiral) closed
string state where the left and right charge +3 operators $\rho_L$ and $\rho_R$ have been
applied on the anti-chiral operators to convert them into charge +2 operators. Charge +2
operator is the (chiral, chiral) state and hence represents
the holomorphic derivative with respect to the corresponding target space modulus. Charge +4
and +2 therefore correspond to the antiholomorphic and holomorphic complex structure moduli, respectively,
of the target space. Recall that here we work in the Type I description,
where the relevant topological twist is the one of B-model. We thus obtain a term similar to
(\ref{deg1}):
\begin{equation}
F^{(0,h_1)}_{\bar{i},\bar{j}}G^{\bar{j} j} D_j F^{(0,h_2)}  ; ~~~~~h_1+h_2 = h\, ,
\end{equation}
where $j$ and $\bar j$ are closed string states.

Finally, charge +3 at a puncture corresponds to the diagonal combination of
(chiral, anti-chiral) plus (anti-chiral, chiral) states which are in fact the K\"ahler
moduli of the target space. These moduli are expected to decouple from the topological
B-model. In the Type I context, they are complexified with
Ramond-Ramond (RR) fields which are associated with continuous shift symmetries in
perturbation theory that make the holomorphic dependence on the K\"ahler moduli trivial.
Charge +3 can also include the identity operator corresponding to exchange of dilaton
in the parent string theory. Indeed, in the example discussed in the section 8, there will
be such a contribution to the holomorphic anomaly from the identity channel in the
closed string degeneration.
\end{enumerate}

Note that for $h=2$ the first type of degeneration is absent, while the second gives the
well known holomorphic anomaly equation for the gauge couplings.
The terms arising from the degeneration with closed string
intermediate states need to be understood further, however in the following we will focus on
the terms coming from the first two types of degenerations that involve open string
intermediate states. The new terms that appear in the holomorphic anomaly equations are
$F^{(0,h)}_n$'s, that is $F^{(0,h)}_{\bar{i}_1...\bar{i}_n; \bar{j}_1...\bar{j}_n}$ where
$\bar{i}$ indices refer to open string insertions of anti-chiral operators carrying charge -1
while $\bar{j}$ indices refer to anti-chiral operators carrying charge +2 (by the action
of $\rho$). In heterotic theory, it was shown that these quantities originate from F-terms
of the form $({\rm Tr}W^2)^h \Pi^n$~\cite{agnth}.
In components they include terms
like ${\cal F}^2 (\lambda^2)^{h-1} \prod_{k=1}^n (\psi_{\bar{i}_k}.\psi_{\bar{j}_k})$ among others,
where $\lambda$ are the gauginos and $\psi$ are non-chiral
matter fermions. The proof of this statement in the case of open string is identical to the
one given for the heterotic case, since in the open string, by going to the double cover, the
spin structure sum boils down to just one sector, as in the heterotic string.
When supersymmetry is broken by a D-term expectation value, the one-loop term
$\Pi{\rm Tr}W^2$ gives rise to fermion masses similar to a ``Higgs" $\mu$-term
in the effective field theory. These masses will be studied in section 7.

\section{$F^{(0,3)}$ in Type I with Magnetized D9 Branes}

In this section, we consider an explicit example of Type I string compactified on a
factorized six-torus $T^6=(T^2)^3$, with magnetized D9 branes. We will then compute the
topological partition function $F^{(0,3)}$ on a world-sheet with three boundaries, in a
$N=1$ supersymmetric configuration corresponding to an appropriate choice of the magnetic
fields. For this purpose, we first recall the main properties of the relevant Riemann
surface $(g=0,h=3)$.

\subsection{Properties of the $(g=0, h=3)$ Riemann surface}

The $(g=0, h=3)$ surface $\Sigma_{(0,3)}$
can be obtained from the double torus of genus 2, by applying
the world-sheet involution (\ref{invh}) exchanging left and right
movers, as described in section~\ref{hsurface} and shown in Fig.~5.
\begin{figure}[h]\hfill
\includegraphics[scale=.6]{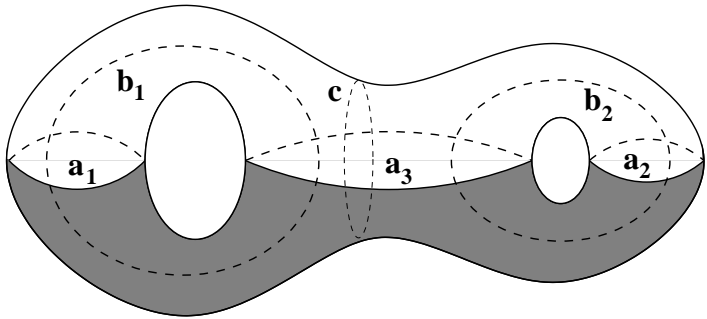}\hfill\hfill
\caption{The $(g=0, h=3)$ surface $\Sigma_{(0,3)}$ is in the upper half of its $g=2$
double-cover. ${\bf c}$ is the dividing geodesics.}
\label{topf4}
\end{figure}
The period matrix $\tau$, invariant under the involution,
is purely imaginary:
\be
\tau=i\left(\begin{matrix}
t_{11} & t_{12} \cr t_{12} & t_{22}\cr
\end{matrix}\right)\, ,
\label{periodmatrix}
\ee
where $t_{ij}$ are three real parameters, dual to the sizes
of the three holes. They correspond to the proper time variables
of the closed string propagation channels.

The ``relative modular group", defined by the modular transformations
$Sp(4,\mathbb{Z})$ that preserve the involution (\ref{invh}), is the
group $GL(2,\mathbb{Z})$~\cite{modular}. It transforms
the period matrix as:
\be
\tau\to D\, \tau\, D^T\, ,
\label{modgroup}
\ee
where $D$ is an arbitrary $2\times 2$ matrix with an inverse of
integer entries. This group, however, in general transforms the boundaries
(fixed under the involution) to linear combinations of the boundaries.
The relevant modular transformations are the ones that act at most as
permutations of the boundaries. This group is generated by
\be
D_1=\left(
\begin{matrix} 0 & 1\cr 1 & 0\cr \end{matrix}
\right)\qquad D_2=\left(
\begin{matrix} -1 & -1\cr 0 & 1\cr \end{matrix}
\right)
\label{generators}
\ee
corresponding to
\be
D_1: t_{11}\leftrightarrow t_{22}\qquad
D_2: \left(\begin{matrix} t_{11}\to t_{11}+t_{22}+2t_{12}\cr
t_{12}\to -t_{12}-t_{22}\end{matrix}\right)\, .
\label{generomega}
\ee
It is easy to see that $D_1^2= D_2^2= (D_1D_2)^3=1$.

Using the above transformations and the positivity of the period matrix,
one can choose as fundamental domain of integration the ordering:
\be
-{\sqrt{t_{11}t_{22}}} \le t_{12}\le 0\le t_{11}\le t_{22}<\infty\, .
\label{order}
\ee
The three degeneration limits, described in section 4, correspond to
the following boundaries of the fundamental domain:
\begin{enumerate}
\item $t_{12}\to 0$, corresponding to shrinking
the dividing geodesics of genus 2. Then $\Sigma_{(0,3)}$
degenerates to a product of two annuli with one common boundary
stretched in the two annuli through a massless open string.
Moreover, the period matrix becomes diagonal with $t_{11}$
and $t_{22}$ the closed string proper times of the two annuli.
\item $t_{12}\to -{\sqrt{t_{11}t_{22}}}$, implying the vanishing
of the period matrix determinant. In this limit, $\Sigma_{(0,3)}$
degenerates into an annulus with a massless open string attached
at one of its boundaries. It amounts to taking the infrared limit of one of
the two gauge loops in the effective field theory.
Then $t_{22}$ is the proper time of the annulus in the closed string
channel while $t_{11}$ parameterizes the length of the open string.
\item $t_{22}\to\infty$, corresponding to pinching an intermediate
closed string connecting an annulus with a disk.
Then $t_{11}$ becomes the proper time of the annulus in the
closed string channel and $t_{12}$ parameterizes the position
of the shrinking hole.
\end{enumerate}

\subsection{Partition functions}

We now discuss the bosonic and fermionic partition functions.
Quantum determinants can be obtained by taking the appropriate square
root of the corresponding expression on the genus 2 double cover.
In the bosonic case, there is a correction factor $R_\Sigma$ that
depends on the involution and on the boundary conditions~\cite{modular}.
It is given by Eq.(\ref{corfin}), as described in section~{\ref{hsurface}.
In the case of a compact boson, one has to multiply the quantum determinant
with the lattice sum of momenta or winding modes (\ref{actions}).
The resulting partition function
for N boundary conditions reads:
\be
Z_B=Z_1^{-1/2}Z^{N,D}_{\rm cl}\qquad
Z^{N}_{\rm cl}=R\sum_{\vec n}e^{-\displaystyle{\pi R^2\over 2}
{\vec n}t{\vec n}^T}\, ,
\label{Zb}
\ee
where ${\vec n}=(n_1,n_2)$ are the winding numbers
around the two $\bf a$ cycles.
In the degeneration limit $t_{22}\to\infty$, non-vanishing $n_2$
windings are exponentially suppressed, and for $n_2=0$ one
recovers the annulus partition function depending on the
closed string proper time $t_{11}$. The $t_{12}$
dependence, corresponding to the position of the shrinking
boundary, drops.
Similarly, in the case of D boundary conditions, one has:
\be
Z^{D}_{\rm cl}=R(\det t)^{-1/2}\sum_{\vec m}
e^{-\displaystyle 2\pi R^2{\vec m} t^{-1}{\vec m}^T}\, .
\label{ZbD}
\ee

The method of section~\ref{hsurface} can also be
applied to the fermionic determinants
giving rise to theta functions. The partition function of a complex fermion
depends on 16 spin structures corresponding to the four boundary
conditions ${\vec a}=(a_1,a_2)$ and ${\vec b}=(b_1,b_2)$ around
the non-trivial cycles ${\bf a}_i$ and ${\bf b}_i$:
\ba
\label{Zf}
Z_f\left[{{\vec a}\atop{\vec b}}\right] &=& Z_1^{-1/2}
\Theta\left[{{\vec a}\atop{\vec b}}\right](t )\\ &=&
Z_1^{-1/2}\sum_{{\vec n}=(n_1,n_2)}
e^{\displaystyle -\pi ({\vec n}+{\vec a})t ({\vec n}+{\vec a})^T
+2i\pi ({\vec n}+{\vec a}){\vec b}^T}\nonumber\, .
\ea
The partition function involves a sum over spin structures with
appropriate coefficients $c\left[{{\vec a}\atop{\vec b}}\right]$,
determined at one loop level. As usually, at higher loops the
corresponding coefficients are determined by the factorization
properties of the vacuum amplitude. Indeed, by considering
the factorization limit $t_{12}\to 0$, we find:
\be
c\left[{{\vec a}\atop{\vec b}}\right]=c_A\left({a_1\atop b_1}\right)\
c_A\left({a_2\atop b_2}\right)\, ,
\label{coefs}
\ee
where $c_A$ denote the corresponding coefficients in the
annulus amplitude.

\subsection{The topological amplitude}

We consider now a toroidal compactification of Type I string theory on three
factorized tori, $T^6=\prod_{I=3,4,5}T^2_I$, and a
$N=1$ supersymmetric configuration
of magnetized D9 branes. Since the string diagram has
three boundaries, we consider in general three brane
stacks associated to the gauge group $U(N_a)$, $a=1,2,3$.
In each of the three abelian factors, there is an internal
magnetic field $H_a^I$ with components along the three
factorized tori $T^2_I$, $I=3,4,5$.  They are quantized
in units of the corresponding areas ${\sqrt G}_I$,
with $G_I$ the determinant of the $T^2_I$ metric,
according to the Dirac quantization condition
$H_a^I=q_a^I/p_a^I{\sqrt G}_I$, where $q_a^I$ is the
respective magnetic flux and $p_a^I$ the wrapping
number of the $a$-th brane around the $I$-th 2-torus.
Note that for each $I$ and $a$, the two integers $p_a^I$
and $q_a^I$ are relatively coprime.
By T-dualizing three directions, one from each $T^2$, one
obtains an equivalent description as a Type IIA orientifold
with D6 branes at angles related to the magnetic fields~\cite{Bach}.
More precisely the angle of each brane relative, for
instance, to the horizontal axis of $T^2_I$ is
$\theta^a_I=\arctan H_a^I\alpha'$. In this representation,
the condition for having $N=1$ unbroken supersymmetry
on the $a$-th stack takes the simple form that the sum
of the corresponding angles in the three tori should vanish:
$\theta^a_3+\theta^a_4+\theta^a_5=0$.

Let us compute now the topological partition function
$F^{(0,3)}$ given by the physical amplitude involving two
gauge fields on one of the 3 boundaries and two gauginos
on another, associated to the effective F-term interaction
$({\rm Tr}W^2)^2$. From the above discussion, it is clear
that the computation is just a particular case of the
general orbifold compactification described in section 3. Indeed,
all twists around {\bf a} cycles are trivial, while
the angles $\theta^a_I$ related to the magnetic fields
play exactly the role of the orbifold twists around the
$\bf b$ cycles, as mentioned already in section 3.
By identifying the first two
boundaries $a=1,2$ with the two ${\bf a}_i$ cycles,
$i=1,2$, of the homology basis of the genus 2 Riemann
surface, and the third boundary $a=3$ with the middle
``horizontal'' cycle, see Fig. 5, one has the relations:
\be
2\pi h_I^1=2(\theta^1_I-\theta^3_I)\qquad ;\qquad
2\pi h_I^2=2(\theta^3_I-\theta^2_I)\, ,
\label{twistangle}
\ee
where $h_I^i$ are the orbifold twists around the two
${\bf b}_i$ cycles.\footnote{
Note that in the presence of orientifold planes, there
are additional diagrams where one or two boundaries
are replaced by crosscaps. Their inclusion is straightforward
and do not change the physical implications of our results.}

Using the identification (\ref{twistangle}),
the calculation goes along the lines of section 3 and the result is given by (\ref{ftoph}).
In section 3, we had not specified the lattice sum involved
explicitly. In the Appendix,
we give the detailed derivation of the lattice sums appearing in $F^{(0,h)}$ in the
example of magnetized D9 branes. Here we just summarize the result for $h=3$. In the T-dual
version, the three stacks of D6 branes on the three boundaries, will be parallel to some primitive
lattice vectors in each plane, $\vec{v}^I_a$, where $a$ labels the three boundaries and $\vec{v}^I$ are
the two dimensional lattice vectors in the $I$-th plane. Since we are
considering factorized torii,
it is sufficient to focus on one plane, therefore in the following we will drop the index $I$
labeling the different planes. In the final formulae we will reinstate the index $I$.
In terms of the magnetic fluxes in the
original D9 branes, these vectors are $\vec{v}_a = (p_a R_1, q_a R_2)$ where $R_1$
and $R_2$ are the two radii in that plane in the T-dual theory
with intersecting branes. The world-sheet can therefore carry windings $n_a \vec{v}_a$ on the $a$-th boundary with integer $n_a$. However they satisfy the constraint that the sum of the windings over the three boundaries
must vanish since it gives the winding on a homologically trivial cycle:
\be
\sum_{a=1}^3 n_a \vec{v}_a = 0\, .
\label{nacons}
\ee
Since we are excluding the case when the three stacks of branes are parallel to each
other in one or more planes (otherwise there would be enhanced supersymmetry),
only one of the
$n_a$'s is independent and it spans a sublattice of integers that depends on the
magnetic flux data $p_a$ and $q_a$. Thus we have a one-dimensional sublattice sum
labeled by, say, $n_1$.

On the world-sheet there is another set of integers which appears because the position of
different stacks of branes is defined only modulo transverse lattice vectors. Since not
all the branes are parallel, we can choose the intersection of two stacks of branes as the origin
of the plane. Then the freedom is only in choosing the transverse position of the third stack of
branes. Thus we have again a one dimensional lattice sum. Upon Poisson resummation over this lattice
we will get a lattice of momenta along the Dirichlet directions of the 3 stacks of branes.
The details are given in the Appendix, but here we just give a simple argument to determine what this
lattice sum would be. Let $*\vec{v}_a$ be the two dimensional
dual of $\vec{v}_a$ and let ${\sqrt G}$ be the
area of the torus (i.e. $R_1R_2$). Then $*\vec{v}_a/{\sqrt G}$ is the primitive vector in the intersection
of the dual momentum lattice and the Dirichlet direction to the $a$-th stack of branes. Thus
the boundary state of the $a$-th branes would carry
momentum vectors $k_a *\vec{v}_a/{\sqrt G}$ with $k_a$
being integers. Conservation of the total momentum gives the same constraint as (\ref{nacons}) with
$n_a$ replaced by $k_a$.
\be
\sum_{a=1}^3 k_a \vec{v}_a =0\, .
\label{kacons}
\ee
This results in again a one dimensional sublattice sum labeled by say $k_1$.

The modular group $SL(2,\mathbb{Z})$, associated to the K\"ahler modulus of $T^2$,
acts on the pair of integers $(n_1, k_1)$ in the usual way.
Since $(n_1,k_1)$ span
only a sublattice of integers subject to the constraint (\ref{nacons}) and (\ref{kacons}),
one might wonder if only a
subgroup of the full $SL(2,\mathbb{Z})$ survives. However these two
constraints are invariant under the full
$SL(2,\mathbb{Z})$ symmetry which implies that the symmetry group is indeed the full $SL(2,\mathbb{Z})$.

To proceed further we need to write a classical solution (before the Poisson resummation) carrying
the above winding numbers and the transverse positions. In terms of the complex coordinate of the plane
$Z= X_1+iX_2$, the boundary conditions imply that $Z$ is untwisted along
the ${\bf a}$ cycles and twisted by
say $g_1=e^{2i\pi h^1}$ and $g_2=e^{2i\pi h^2}$ along the ${\bf b}_1$ and ${\bf b}_2$
cycles of the genus 2 double cover of the surface
$\Sigma_{(0,3)}$. Denoting by $\{g\}$ the collection of the twists
$g_1$ and $g_2$, we note that
by Riemann-Roch theorem there is only one linearly independent holomorphic twisted differential
(Prym Differential) $\omega_{\{g\}}$. To write the classical solution we need the
twisted holomorphic differentials $\omega_{\{g\}}$, $\omega_{\{g^{-1}\}}$ and their complex
conjugates $\bar{\omega}_{\{g\}}$, $\bar{\omega}_{\{g^{-1}\}}$. The solution is of the form
\be
Z(P) = L \int_{P_0}^P \omega_{\{g\}} + \tilde{L} \int_{P_0}^P \bar{\omega}_{\{g^{-1}\}}\, ,
\ee
where $L$ and $\tilde{L}$ are determined in terms of winding number and transverse position
data and the normalization condition for the twisted differential. Since ${\bf a}$ cycles are
untwisted, we can choose the normalization condition (for convenience)
\be
\int_{{\bf a}_1} \omega_{\{g\}} = g_1^{1/2}\, .\label{omnorm}
\ee

Note that the integral around $a_2$ cycle is not independent since integrating over the trivial
cycle $\prod_{i=1}^2 ({\bf a}_i {\bf b}_i {\bf a}_i^{-1} {\bf b}_i^{-1})$
gives the constraint \be [(1-g_1)\int_{{\bf a}_1}+
(1-g_2)\int_{{\bf a}_2}] \omega_{\{g\}} =0.\label{omnorm1}\ee
By using Fig. 6 (for $h=3$), one can evaluate
\be
\int_{\Sigma_{(0,3)}} \omega_{\{g\}} \wedge \bar{\omega}_{\{g\}} =
\tau_{\{g\}}-\bar{\tau}_{\{g\}} ~~~;~~~~ \tau_{\{g\}}= D_{\{g\}} +
\frac{1}{2}\frac{(1-g_1)(1-g_2 g_1^{-1})}{1-g_2}\, ,
\label{tautwist}
\ee
where
\be
D_{\{g\}} = 2i\frac{(g_1 g_2)^{-\frac{1}{2}}}{1-g_2^{-1}}
\int_{{\bf b}_1 {\bf b}_2 {\bf b}_1^{-1} {\bf b}_2^{-1}} \omega_{\{g\}}\, .
\label{Dtwist}
\ee
Note that the individual cycles ${\bf b}_i$ are not closed due to the twists, but the cycle
${\bf b}_1{\bf b}_2 {\bf b}_1^{-1} {\bf b}_2^{-1}$ is closed.
Due to the symmetry of the double cover under the
involution (\ref{invh}), $D$ turns out to be purely imaginary. By using the identity
$\int_{\Sigma_{(0,3)}} \omega_{\{g\}} \wedge {\omega}_{\{g^{-1}\}}=0$
we have the relation $\tau_{\{g\}}=\tau_{\{g^{-1}\}}$.

The final result of the
lattice contribution as shown in the Appendix (including all the three planes) is\footnote{Here and in the Appendix [from Eq.(A.25) onwards] we do not keep track of overall factors that are completely moduli- and
flux data-independent.}
\be
Z_{\rm lattice}= \prod_I p_I\left({{\rm Im}\tilde{\tau}_{\{g_I\}}\over{\rm Im}T}\right)^{1/2}
\sum_{(n^I_1,k^I_1)} e^{2i\pi( n^I_1 \alpha^I_1 +
k^I_1 \alpha^I_2)}
e^{\frac{i\pi}{{\rm Im}T_I} |n^I_1 T_I + k^I_1|^2\tilde{\tau}_{\{g_I\}}}\, ,
\label{zlattice03}
\ee
where $T_I=B_I+i{\sqrt G}_I$ is the usual K\"ahler modulus of the torus $T^2_I$,
with $B_I$ the two-index antisymmetric tensor. The parameter
$\tilde{\tau}_{\{g_I\}} = \frac{|v^I_1|^2}{{\rm Im}T_I}
\tau_{\{g_I\}}$ is independent of the modulus
$T_I$, although it depends on the world-sheet moduli and the flux data.
Furthermore, $\alpha_1$ is the effective Wilson
line along the world-volume of the branes and $\alpha_2$ is the effective transverse position of
the branes (which is T-dual to the second component of the Wilson line in the D9 brane theory).
The integer $p_I$ is the smallest positive integer such that
\be
p_I \frac{v_3^I.*v_1^I}{v_3^I.*v_2^I} \in \mathbb{Z}.
\label{pIcon}
\ee
 Note that the above equation implies
 \be p_I \frac{v_2^I.*v_1^I}{v_3^I.*v_2^I}
 \in \mathbb{Z}.\ee

{}From the constraints (\ref{nacons}) and (\ref{kacons}) it
 follows that $n_1^I$ and $k_1^I$ are arbitrary integer multiples of $p_I$.
As observed
in subsection 5.1 [preceding Eq.(\ref{generomega})], the relevant modular group is the permutation
group $S_3$ that permutes the three boundaries. Since we have treated the three boundaries asymmetrically
in arriving to Eq.(\ref{zlattice03}), it is not manifest that the result is modular invariant. For example, we have normalized
the Prym differential along the first boundary and the lattice momenta appearing in the expression
refer directly to the boundary state of the brane attached to this boundary.
Indeed, under the permutation of the boundaries, $\tau_{\{g_I\}}$, $p_I$ and $|v_1^I|$ transform nontrivially. However, the combination $p_I |v_1^I| \tau_{\{g_I\}}^{\frac{1}{2}}$ is invariant. For instance, under the exchange of first and second boundaries, \be
v_1^I \to v_2^I,\qquad p_I \to p_I \left|\frac{v_3^I.*v_2^I}{v_3^I.*v_1^I}\right|\quad ~\makebox{and}~ \quad \tau_{\{g_I\}} \to \left|\frac{1-g_1^I} {1-g_2^I}\right|^2 \tau_{\{g_I\}}.\ee
The latter can be seen by the fact that under the exchange of ${\bf a}_1$ and ${\bf a}_2$ cycles, $\omega_{\{g_I\}}$ [normalized by Eq.(\ref{omnorm}) around ${\bf a}_1$ cycle],  is transformed to $-\frac{1-g_1^I} {1-g_2^I}\omega_{\{g_I\}}$, where we used the constraint (\ref{omnorm1}). The statement that $p_I |v_1^I| \tau_{\{g_I\}}^{\frac{1}{2}}$ is invariant now follows from the
relation \be \left|\frac{1-g_1^I}{1-g_2^I}\right|=\left|\frac{v_3^I.*v_1^I}{v_3^I.*v_2^I}\right| \frac{|v_2^I|}{|v_1^I|}.\ee As a result, $Z_{\rm lattice}$ as given in (\ref{zlattice03}) is invariant under the permutation group.

In the topological partition function, we have also the insertion of $G^-$'s which are folded
with the Beltrami differentials.
As mentioned earlier, each $\bar{\psi}^I$ field,
being of dimension 1 in the topological theory,  has now only one
zero mode (for each plane $I$).
It will therefore be replaced by $\omega_{\{g_I^{-1}\}}$. Similarly $\partial Z_I$ will be
replaced by the zero mode
\be
\partial Z_I =  \frac{v_I}{{\rm Im}T_I}(n^I_1 T_I + k^I_1)\omega_{\{g_I\}}\, .
\ee
Integrating the fermion zero modes we get
\be
\prod_{a=1}^3 \int \mu_a G^- =    \prod_{I=1}^3  \frac{v_I}{{\rm Im}T_I}(n^I_1 T_I + k^I_1)
\det \int \mu_a  h_J \quad ; \quad h_J =
\omega_{\{g_J\}} \omega_{\{g_J^{-1}\}}\, .
\label{measure}
\ee
We can further evaluate the determinant above by noting that under the deformation of the
complex structure represented by the Beltrami differential $\mu_a$, the twisted $(1,0)$
form  $\omega_{\{g_I\}}$ picks up a $(0,1)$ form given by
\be
\omega_{\{g_I\}}  \rightarrow \omega_{\{g_I\}} + C \bar{\omega}_{\{g^{-1}_I\}} \quad ; \quad  C
\bar{\omega}_{\{g^{-1}_I\}} = \mu_a \omega_{\{g_I\}}\ ~~{\rm modulo\ an\ exact\ form}\, .
\label{cdef}
\ee
As a result
\ba
\int \mu_a \omega_{\{g_I\}} \omega_{\{g^{-1}_I\}} &=&  C \int  \bar{\omega}_{\{g^{-1}_I\}}
\omega_{\{g^{-1}_I\}} \nonumber \\&=& C(\tau_{\{g_I\}}-\bar{\tau}_{\{g_I\}}).
\ea
Now we will relate this to the variation of the twisted $\tau_{\{g_I\}}$.
The moduli dependent part of the
latter, as seen from Eq.(\ref{tautwist}) and (\ref{Dtwist}),
is proportional to the ratio of the periods
$\int_{{\bf b}_1{\bf b}_2{\bf b}_1^{-1}{\bf b}_2^{-1}}\omega_{\{g_I\}}
/\int_{{\bf a}_1} \omega_{\{g_I\}}$. From the variation of the twisted differential given in
(\ref{cdef}) we find
\begin{eqnarray}
\delta_a \tau_{\{g_I\}} &=& C(\bar{D}_{\{g_I^{-1}\}}-D_{\{g_I\}})
\nonumber \\ &=& C(\tau_{\{g_I\}}-\bar{\tau}_{\{g_I\}}) \nonumber \\ &=&
\int \mu_a \omega_{\{g_I\}} \omega_{\{g^{-1}_I\}}\, ,
\end{eqnarray}
where in the second equality we have used Eqs.(\ref{tautwist}) and (\ref{Dtwist}) and
the fact that $\tau_{\{g_I\}}=\tau_{\{g_I^{-1}\}}$.
The determinant appearing in (\ref{measure}) therefore just gives the Jacobian of the
transformation from the moduli to $\tau_{\{g_I\}}$. If the latter, for $I=3,4,5$ are independent
functions on the moduli space, then the Jacobian is non-vanishing  and the resulting measure
of integration on the moduli space becomes $\prod_I d\tau_{\{g_I\}}$.

In the topological twisted theory the non-zero modes of the bosons and fermions cancel leaving
only the correction factor coming from the boundary conditions on the bosons discussed in section 2
and summarized in Eq.(\ref{corfin}). For the twisted bosons, $\tau$ in (\ref{corfin}) is replaced by
the corresponding twisted $\tau$. The resulting ${\rm det}\tau$ factor cancels the one appearing in
lattice partition function (\ref{zlattice03}).

The above calculations were done in the D6 brane theory. By T-dualizing we can go to the magnetized
D9 brane theory. This amounts to replacing $T_I \to U_I$ and ${\rm Im}U_I \to {\rm Im}T_I =
\sqrt{G_I}$. Here, $U$ is the usual complex structure modulus of the torus $T^2$,
given in terms of its metric $G_{ij}$, $i,j=1,2$: $U=(G_{12}+i{\sqrt G})/G_{11}$.
The resulting topological partition function becomes:
\be
F^{(0,3)}=3N\int \prod_{I=3}^5 d\tilde{\tau}_{\{g_I\}} p^I 
f\left[U_I, \vec{\alpha}_I;-i\tilde{\tau}_{\{g_I\}}\right]\, ,
\label{F03}
\ee
where the integration variables
\be
\tilde{\tau}_{\{g_I\}}= \sqrt{G_I} (p_1^I)^2 \left[1+(H^I_1)^2\right] \tau_{\{g_I\}}\, .
\label{F03int}
\ee
Here $H^I_a=\frac{q^I_a}{p^I_a\sqrt{G_I}}$ is  the magnetic field on the $a$-th brane stack in
the $I$-th plane. The function $f$ is given by:
\be
f(U,\vec{\alpha},l)= \frac{1}{{\rm Im} U}
\sum_{n_1,n_2} (n_1+n_2\bar{U})
e^{2i\pi\vec{n}\cdot \vec{\alpha}}
e^{-{\pi\over {\rm Im}U}|n_1+n_2 U|^2l}\, ,
\label{F03f}
\ee
where the lattice sum of integers $\vec{n}^I$ for each $I$ satisfy
the conditions
\be
p^I_1 \vec{n}_I + p^I_2 \vec{n}'_I + p^I_3 \vec{n}''_I=
q^I_1 \vec{n}_I + q^I_2 \vec{n}'_I + q^I_3 \vec{n}''_I=0
\label{npqconst}
\ee
for some integer vectors $\vec{n}'_I$ and $\vec{n}''_I$.
$\vec{\alpha}_I$ is the Wilson line in the $I$-th plane
and is given in terms of a certain linear combination of the Wilson lines on the three brane stacks:
\be
\vec{\alpha}^I= \vec{\alpha}^I_1 - \frac{p^I_3 q^I_1-q^I_3 p^I_1}{p^I_3 q^I_2-q^I_3 p^I_2}\vec{\alpha}^I_2
+\frac{p^I_2 q^I_1-q^I_2 p^I_1}{p^I_3 q^I_2-q^I_3 p^I_2}\vec{\alpha}^I_3\, ,
\label{effectivewilson}
\ee
where the subscripts $1,2,3$ refer to the three different stacks of branes on the three boundaries
of the world-sheet. Obviously,
the result vanishes for ${\vec{\alpha}_I}=0$ for any $I$, due to the
${\vec n}_I\to -{\vec n}_I$ symmetry of the lattice sum.
Note that non-trivial $\vec \alpha_I$, although
breaks the corresponding R-symmetry, does not change the spectrum.
The positive integers $p_I$ appearing in (\ref{F03}) satisfy (\ref{pIcon})
 which can be reexpressed in terms
of the flux data as $p_I$ being the smallest positive integer such that
\be
p_I\frac{p^I_3 q^I_1-q^I_3 p^I_1}{p^I_3 q^I_2-q^I_3 p^I_2} \in \mathbb{Z} ~~~
\Rightarrow ~~~ p_I
\frac{p^I_2 q^I_1-q^I_2 p^I_1}{p^I_3 q^I_2-q^I_3 p^I_2} \in \mathbb{Z}\, .
\label{pi}
\ee

It is worth making some comments on the topological partition function (\ref{F03}).
The first comment is regarding modular invariance. From the constraint
(\ref{npqconst}) it follows that $n_1^I$ and $n_2^I$ are arbitrary integer
multiples of $p_I$. The discussion following Eq.(\ref{pIcon}) then implies
that (\ref{F03}) is invariant under the permutation of the boundaries.

The second comment is regarding the target space duality properties of (\ref{F03}). We have already
mentioned that it has full $SL(2,\mathbb{Z})_U$ symmetry despite the fact that the sum is over a sublattice
defined by the constraints (\ref{npqconst}). This is because the constraints are $SL(2,\mathbb{Z})_U$ covariant.
As for the monodromy under shifts of the Wilson lines we note that
the constraints imply that ${\vec n}_I \in p^I {\vec Z}_I$ where ${\vec Z}$
is a two-dimensional vector with arbitrary integer components. It follows then that $F^{(0,3)}$ is
invariant under shifts $\vec{\alpha}_I \to \vec{\alpha}_I + \vec{Z}_I/p_I$.

The topological partition function (\ref{F03}) appears deceptively simple as products of
three separate integrals, however, the domain of integration over ${\tilde\tau}_{\{g_I\}}$ is very
complicated and in general mixes the three integrals.
There are some comments worth making about $F^{(0,3)}$:
\begin{itemize}
\item Taking derivatives with respect to anti-holomorphic moduli should result in
a total derivative in the world-sheet moduli space, as follows from the general discussion of
section 4.
The anti-holomorphic moduli in the present case are the complex structure closed string moduli
$\bar{U}_I$ and the complexified Wilson lines
\be
A_I=  \alpha_2^I - U_I \alpha_1^I\, .
\ee
The physical quantities are invariant under $A_I \rightarrow A_I+1/p_I$ and $A_I \rightarrow
A_I+U_I/p_I$.
The derivatives with respect to the anti-holomorphic moduli can be easily evaluated
using the identities:
\ba
{\partial f\over\partial{\bar A}} &=& {1\over{\rm Im}U}{\partial Z \over \partial l}\quad ;\quad
Z={\sum_{n_1,n_2}} e^{2i\pi{\vec a}\cdot{\vec n}}
e^{-{\pi\over {\rm Im}U}|n_1+n_2 U|^2l}\nonumber\\
{\partial f\over\partial{\bar U}} &=& -{1\over 4{\rm Im}U}{\partial{(lf)}\over\partial l}\
-\frac{{\rm Im}A}{({\rm Im} U)^2}\frac{\partial Z}{\partial l} ,
\label{ident}
\ea
which follow from the form of the function $f$ in Eq.(\ref{F03f}).

\item The dependence on the K\"ahler moduli of the tori, namely $\sqrt{G_I}$, should drop out
because the complexification of these moduli involve the Ra\-mond-Ramond $B$ fields. Since
the latter are associated with continuous shift symmetries in perturbation theory, the
dependence on them should be trivial.
\end{itemize}

\section{Gaugino Masses}

In this section, we compute the Majorana gaugino masses
in the general non-supersymmetric case with the sum of
the brane angles different from zero. Using their relation to the
orbifold twists (\ref{twistangle}), we parameterize the
supersymmetry breaking as
\be
\sum_{I=3}^5 h_I=-2\epsilon\, ,
\label{Dterm}
\ee
where for simplicity we dropped the cycle indices.
The parameter $\epsilon$ fixes the scalar masses and
in the weak field limit $\epsilon\to 0$, it is proportional
to the expectation value of the corresponding abelian
D-term. In this limit, the gaugino masses are determined
from the effective F-term $({\rm Tr}W^2)^2$ and should therefore
be identified with the topological partition function $F^{(0,3)}$.

The gaugino vertices in the $-1/2$ ghost picture are
given in Eq.(\ref{lvertex}) and are inserted on one of the three
boundaries of the $(g=0,h=3)$ Riemann surface.
Moreover one has to insert three picture changing operators
$e^\varphi T_F$, which we also choose to be on the
boundaries. One of them is needed to change the
ghost picture of one gaugino to $+1/2$, while the other
two arise from the integration over the supermoduli:
\be
m_{1/2}=g_s^2\int [dt]\int dxdy\ {\cal A}\quad ;\quad
{\cal A}=
\left\langle V^{(-1/2)}_+(x)V^{(-1/2)}_-(y)\prod_{i=1}^3
e^\varphi T_F(z_i)\right\rangle\, ,
\label{amp}
\ee
where $g_s$ is the string coupling and its power takes into
account the normalization of the gaugino kinetic terms
on the disk. The moduli integration is over the fundamental
domain (\ref{order}), while $x$ and $y$ are integrated
over the gaugino boundary. The dependence on the
positions $z_i$ of the picture changing operators is a gauge
artifact and should disappear from the physical amplitude.

By internal charge conservation, since both gauginos carry
charge $+1/2$ for $\phi_3$, $\phi_4$ and $\phi_5$, only the
internal part of the world-sheet supercurrents (\ref{tfint}) 
contributes; each $T_F^{int}$ should provide $-1$ charge for
$\phi_3$, $\phi_4$ and $\phi_5$, respectively.
The amplitude (\ref{amp}) then becomes:
\ba
{\cal A} &=& \langle e^{-\varphi/2}(x)e^{-\varphi/2}(y)
\prod_{I=3}^5\! e^\varphi(u_I)\rangle
\prod_{I=1}^2\! \langle e^{i\phi_I/2}(x)e^{-i\phi_I/2}(y)\rangle
\nonumber\\ & &\times
\prod_{I=3}^5\! \langle e^{i\phi_I/2}(x)e^{i\phi_I/2}(y)e^{-i\phi_I}(u_I)
\rangle\, ,
\label{ampA}
\ea
where $\{ u_I\}$, $I=3,4,5$, is a permutation  of $\{ z_i\}$,
$i=1,2,3$, and an implicit summation over all permutations
is understood.
Performing the contractions for a given spin structure $s$, one finds:
\ba
{\cal A}_s &=&
{\theta_s^2({1\over 2}(x-y))\prod_{I=3}^5\theta_{s,-h_I}
({1\over 2}(x+y)-u_I)\,\partial X_{h_I}(u_I)\over
\theta_s({1\over 2}(x+y)-\sum_{I=3}^5u_I+2\Delta)}\nonumber\\
&&\times{\sigma(x)\sigma(y)\over\prod_{I<J}^{3,4,5}E(u_I,u_J)
\prod_{I=3}^5\sigma^2(u_I)}\times{Z_2\over Z_1^4
\prod_{I=3}^5Z_{1,h_I}} Z_{lat}\, ,
\label{ampAs}
\ea
where $\theta_s$ is the genus-two theta-function of spin structure $s$,
$E$ is the prime form, $\sigma$ is a one-differential with no zeros or
poles and $\Delta$ is the Riemann $\theta$-constant.
$Z_{1,h_I}$ is the (chiral) determinant of the
$h_I$-twisted $(1,0)$ system, and
$Z_2$  is the chiral non-zero mode determinant of the $(2,-1)$
{\it b-c} ghost system. Finally, $Z_{lat}$ stands for all zero-mode
parts of space-time and internal coordinates, while an implicit
summation over lattice momenta should be performed, taking into
account also the $\partial X_I$ factors in (\ref{ampAs}).

In order to perform an explicit sum over spin structures, one
should choose the positions of the picture changing operators
to satisfy a condition that makes the argument of the
$\theta$-function in the denominator equal to $(x-y)/2$,
so that one $\theta_s$ factor simplifies. The resulting
relation however,
\be
\sum_{i=1}^{i=3}z_i=y+2\Delta\, ,
\label{gauged}
\ee
is not allowed. To bypass this difficulty, we insert in the
amplitude (\ref{amp}) another vertex of an open string Wilson
line associated to the $I=3$ internal plane, in the $-1$ ghost picture:
\be
V^{(-1)}=:e^{-\varphi}\psi_3(w):\, ,
\label{extravertex}
\ee
accompanied by a forth picture changing operator at the
boundary position $z_4$. Obviously, we should also perform
a third integration over its position $w$. The new vertex brings
another unit of charge for $\phi_3$, and thus, two of the four
supercurrents should provide $-1$ charge for $\phi_3$.
As we will demonstrate below, one can now choose an appropriate
gauge condition which allows to perform the spin structure
sum and show that the amplitude can be written as the variation
with respect to the Wilson line of the original one (\ref{ampAs}),
evaluated formally using the ``forbidden" gauge choice (\ref{gauged}).

Indeed, the correlators in (\ref{ampA}) now become:
\ba
\label{ampAn}
{\cal A}\!\!\! &=&\!\!\! \langle e^{-\varphi/2}(x)e^{-\varphi/2}(y)e^{-\varphi}(w)
\prod_{I=3}^6 e^\varphi(u_I)\rangle
\prod_{I=1}^2 \langle e^{i\phi_I/2}(x)e^{-i\phi_I/2}(y)\rangle\\
&\times&\!\!\!
\langle e^{i\phi_3/2}(x)e^{i\phi_3/2}(y)e^{i\phi_3}(w)\prod_{I=3,6}
e^{-i\phi_3}(u_I)\rangle
\prod_{I=4,5} \langle e^{i\phi_I/2}(x)e^{i\phi_I/2}(y)e^{-i\phi_I}(u_I)
\rangle\, ,\nonumber
\ea
where $\{ u_I\}$, $I=3,\dots,6$, is a permutation  of $\{ z_i\}$,
$i=1,2,3,4$, and an implicit summation over all permutations
is understood. Performing the contractions, one finds that the
expression (\ref{ampAs}) is modified as:
\ba
{\cal A}_s\!\!\!\! &=&\!\!\!\!
{\theta_s^2({1\over 2}(x-y))\theta_{s,-h_3}({1\over 2}(x+y)-u_3+w-u_6)\
\!\prod_{I=4,5}\theta_{s,-h_I}({1\over 2}(x+y)-u_I)\over
\theta_s({1\over 2}(x+y)+w-\sum_{I=3}^6u_I+2\Delta)}\nonumber\\
& & \times\, {\sigma(x)\sigma(y)\sigma^2(w)
\over\prod_{I<J}^{3,4,5}E(u_I,u_J)\prod_{I=3}^6\sigma^2(u_I)}
{E(w,u_4)E(w,u_5)\over E(u_6,u_4)E(w-u_6,u_5)}\label{ampAsn}\\
& & \times\, {Z_2Z_{lat}\over Z_1^4\prod_{I=3}^5Z_{1,h_I}}
\prod_{I=3,6}\partial X_{3,h_3}(u_I)
\prod_{I=4,5}\partial X_{I,h_I}(u_I)\, .\nonumber
\ea
To perform the spin structure sum, we use the allowed
gauge condition:
\be
\sum_{i=1}^{i=4}z_i=y+w+2\Delta\, .
\label{gaugedn}
\ee
The sum over spin structures converts the first factor in (\ref{ampAsn}) to:
\be
\theta_\epsilon(x-\Delta)\theta_{h_3+\epsilon}(u_3+u_6-w-\Delta)
\theta_{h_4+\epsilon}(u_4-\Delta)\theta_{h_5+\epsilon}(u_5-\Delta)\, ,
\label{spinstr}
\ee
where we used Eq.(\ref{Dterm}).

We now use three bosonization formulae. First,
\ba
\label{bosfor1}
\theta_{h_3+\epsilon}(u_3+u_6-w-\Delta)
{E(u_3,u_6)Z_1^{-1/2}\over E(u_3,w)E(u_6,w)}
{\sigma(u_3)\sigma(u_6)\over\sigma(w)}\\ =
\langle{\overline\psi}_{h_3+\epsilon}(u_3)
{\overline\psi}_{h_3+\epsilon}(u_6)
\psi_{-h_3-\epsilon}(w)\rangle Z_{1,h_3+\epsilon}\, ,
\nonumber
\ea
where $\psi$ and $\overline\psi$ are conformal fields of
dimension zero and one,
respectively, twisted as indicated by their subscripts. Then
\be
\theta_{h_I+\epsilon}(u_I-\Delta)\sigma(u_I)Z_1^{-1/2}=
\omega_{h_I+\epsilon}(u_I)Z_{1,h_I+\epsilon}\, ,
\label{bosfor2}
\ee
where $\omega_h$ is an abelian differential twisted by $h$. Finally,
\be
\theta_\epsilon(\sum_{I=3}^6 u_I-w-3\Delta)
{\prod_{I<J}E(u_I,u_J)\over\prod_{I=3}^6E(u_I,w)}
{\prod_{I=3}^6\sigma^3(u_I)\over\sigma^3(w)}Z_1^{-1/2}\!\!=\!\!
\langle\prod_{I=3}^6b_\epsilon(u_I)c_{-\epsilon}(w)
\rangle Z_{2,\epsilon}, 
\label{bosfor3}
\ee
where the correlator and the non-zero mode determinant
of the {\it b-c} ghost system in the {\em r.h.s.} are twisted
according to the subscripts.

Multiplying the amplitude (\ref{ampAsn}) by
$\theta_\epsilon(y-\Delta)/\theta_\epsilon(\sum_{I=3}^6 u_I-w-3\Delta)$
which is equal to the identity by our gauge condition (\ref{gaugedn}),
and using the spin structure sum (\ref{spinstr}) and the
bosonization formulae (\ref{bosfor1}) - (\ref{bosfor3}), we obtain:
\ba
{\cal A} &=& \theta_\epsilon(x-\Delta)\theta_\epsilon(y-\Delta)
\prod_{I=3}^5\left({Z_{1,h_I+\epsilon}\over Z_{1,h_I}}\right)
{\langle{\overline\psi}_{h_3-\epsilon}(u_3)
{\overline\psi}_{h_3+\epsilon}(u_6)
\psi_{-h_3-\epsilon}(w)\rangle\over
\langle\prod_{I=3}^6b_\epsilon(u_I)c_{-\epsilon}(w)\rangle}
\nonumber\\ &&
\times\prod_{I=4,5}\omega_{h_I+\epsilon}(u_I)
\sigma(x)\sigma(y){Z_2\over Z_{2,\epsilon}}
{1\over Z_1^3} {\prod_{I=3,6}\partial X_{3,h_3}(u_I)
\prod_{I=4,5}\partial X_I(u_I)}.
\label{ampAf}
\ea
Replacing $\omega_{h_I+\epsilon}(u_I)$ by
${\overline\psi}_{h_I+\epsilon}(u_I)$ and using the bosonization
formula (\ref{bosfor2}) twice, one finds:
\ba
{\cal A}={\langle\prod_{i=1}^4G^-_\epsilon(z_i)
\psi_{-h_3-\epsilon}(w)\rangle\over
\langle\prod_{i=1}^4b_\epsilon(z_i)c_{-\epsilon}(w)\rangle}
\prod_{I=3}^5\left({Z_{1,h_I+\epsilon}\over Z_{1,h_I}}\right)
{Z_2\over Z_{2,\epsilon}}
\left({Z_{1,\epsilon}\over Z_1}\right)^2\!\!
\omega_\epsilon(x)\omega_\epsilon(y),
\label{ampAftop}
\ea
where $G^-_\epsilon$ is the internal $N=2$ supercurrent
twisted by $\epsilon$.

One can now show that the ratio of correlation functions
in (\ref{ampAftop}) is independent of the positions $z_i$:
\be
{{\cal N}\over {\cal D}}\equiv
{\langle\prod_{i=1}^4G^-_\epsilon(z_i)
\psi_{-h_3-\epsilon}(w)\rangle\over
\langle\prod_{i=1}^4b_\epsilon(z_i)c_{-\epsilon}(w)\rangle}
=B_\epsilon\partial X_{3,h_3}\omega_{-h_3}(w)\, ,
\label{ident1}
\ee
with $B_\epsilon$ a constant to be determined. Indeed, as
a function of $w$, both numerator ${\cal N}$ and denominator ${\cal D}$
have first order poles when $w\to z_i$ with residues
$\langle\prod_{i=1}^3G^-_\epsilon(z_i)\rangle\partial X_{3,h_3}$
and $\langle\prod_{i=1}^3b_\epsilon(z_i)\rangle$, when for
instance $w\to z_4$. These are equal, up to a $z_i$-independent
multiplicative factor $B_\epsilon$. Consider now the combination
${\cal N}-B_\epsilon\partial X_{3,h_3}\omega_{-h_3}(w){\cal D}$.
This is holomorphic in $w$ twisted 0-form and therefore vanishes
identically. Thus, $B_\epsilon$ is given by:
\be
B_\epsilon\omega_{-h_3}(w)=
{\langle\prod_{i=1}^3G^-_\epsilon(z_i)\rangle\over
\langle\prod_{i=1}^3b_\epsilon(z_i)\rangle}\, .
\label{Bepsilon}
\ee
Note that this is reduced to the same result as the one that would
be obtained using the ``forbidden" gauge condition (\ref{gauged}),
with an additional differentiation with respect  to the Wilson line (\ref{extravertex})
associated to the presence of $\partial X_{3,h_3}\omega_{-h_3}(w)$
in (\ref{ident1}).

It follows that the gaugino mass (\ref{amp}) is given by:
\be
m_{1/2}=g_s^2\int [dt] B_\epsilon
\prod_{I=3}^5\left({Z_{1,h_I+\epsilon}\over Z_{1,h_I}}\right)
{Z_2\over Z_{2,\epsilon}}
\left({Z_{1,\epsilon}\over Z_1}\right)^2\, ,
\label{ampr}
\ee
where we performed the boundary integrals over $x$ and $y$,
using the canonical normalization over the $\bf a$-cycles
$\int_{\bf a}\omega_\epsilon=1$, since $\epsilon$ is a twist
around the $\bf b$-cycles. In the limit of small supersymmetry
breaking scale $\epsilon\to 0$, one has
$Z_{1,\epsilon}\simeq\epsilon Z_1$, and thus
\be
m_{1/2}\simeq g_s^2\epsilon^2\int [dt] B_0=g_s^2\epsilon^2 F^{(0,3)}\, .
\label{m12}
\ee

\section{$\Pi$-terms and Matter  Fermion Masses}

In the previous section, we studied the (Majorana) gaugino mass terms
generated via D-term supersymmetry breaking.
They originate from the supersymmetric F-term $({\rm Tr}W^2)^2$ once the auxiliary
D-component of
the vector superfield acquires a non-zero VEV along a magnetized $U(1)$ group factor.
The corresponding topological coupling is given by $F^{(0,3)}$,
with the three boundaries of the world-sheet attached (in a T-dual picture) to
three different stacks of D6 branes intersecting at
angles in the internal compactification space.
In order to get a non-zero answer, we had to further break the
discrete R-symmetry by turning on
suitable Wilson lines. In the following we analyze
the structure of the terms appearing in the holomorphic anomaly equation for
$F^{(0,3)}$. The are of the form $\Pi{\rm Tr}W^2$, where $\Pi$ is a chiral
projection of a non-holomorphic function of chiral superfields. Generically,
its lower component includes a fermion bilinear of the form
${\bar\Psi}_{\bar i}{\bar\Psi}_{\bar j}$. Hence, upon D-term supersymmetry
breaking, they will also induce some fermion masses, in this case of Dirac type.

In order to examine the holomorphic anomaly of $F^{(0,3)}$, we
take an anti-holomorphic derivative $\partial_{\bar{i}}$,
with respect to an open string Wilson line $\bar{i}$. From the analysis
of section 4, one finds contributions from three possible degeneration
limits. The two open string degenerations, (1) and (2),
involve $F^{(0,2)}_{\bar{i}; \bar{j}}$, where $\bar{j}$ denotes an
intermediate anti-chiral open string state.
In this section, we restrict our discussion to brane configurations that do
not involve parallel stacks in any of the three internal planes. This means
that in the supersymmetric limit there are no $N=2$ supersymmetric sectors
in the one-loop partition function.\footnote{In the next section however, we will consider
examples with such $N=2$ sectors.}
In the absence of such sectors, the dividing degeneration (1) does not contribute
because $\bar{j}$ is an untwisted operator and the corresponding annulus
diagrams vanish. In the handle degeneration limit (2) however,
since the open string propagating through the handle stretches between two
non-parallel brane stacks, it is necessarily twisted. The third degeneration limit (3)
involves intermediate closed strings which, in $T^6$ compactifications, are always
untwisted. Hence, the corresponding annulus diagrams also vanish.

Thus, the only quantity that appears in the holomorphic anomaly of $F^{(0,3)}$
arises from Eq.(\ref{deg2}), and involves $D_j F^{0,2}_{\bar{i}; \bar{j}}$,
where $j$ and $\bar{j}$ label chiral and anti-chiral
twisted open string states with $U(1)$ charges +1 and +2 respectively. These are therefore
bi-fundamental states
and represent open strings stretched between two different stacks of D6 branes
(say $a=1$ and $a=2$) that are not parallel to each other.
If the intersection of these two stacks of D6 branes
preserves supersymmetry then the twist
angles $h_I$ in the three internal planes (tori), $I=3,4,5$, satisfy $\sum_I h_I=1$ and the
corresponding twisted states are massless. In the topological theory
\begin{equation}
D_j F^{(0,2)}_{\bar{i}; \bar{j}} =  \int dt^1 dt^2 \int dx \langle\left\{ \prod_{m=1}^2
\int \mu_m (G_L^-+ G^-_R)\right\} V_j(u) V_{\bar{i}}(x) V_{\bar{j}}(y)\rangle^{\cal T}
\end{equation}
with
\begin{eqnarray}
V^{\cal T}_j &=& : \sigma e^{i\sum_I h_I \phi_I}:\nonumber\\
V^{\cal T}_{\bar{j}} &=& :\bar{\sigma} e^{i\sum_I (1-h_I) \phi_I}: \\
V^{\cal T}_{\bar{i}} &=& :e^{-i\phi_3}:\nonumber
\end{eqnarray}
where, for concreteness, we chose $\bar{i}$ to indicate the
anti-holomorphic derivative with respect to the Wilson line in the $I=3$ plane.
$\sigma$ and $\bar{\sigma}$ denote the
bosonic twisted fields. We have included the superscript ${\cal T}$ to
indicate that these are the vertices
in the topological theory. Note that while $V^{\cal T}_{\bar{i}}$ has twisted $U(1)$ charge $-1$ and
dimension 1 (and hence its position $x$ is integrated on the world-sheet boundary),
the operators $V^{\cal T}_j$ and $ V^{\cal T}_{\bar{j}}$ have dimension 0 and charges +1 and +2 respectively.
The world-sheet moduli space,
labeled by $t^m$, with the corresponding Beltrami differentials $\mu_m$,
is two dimensional, one being the usual modulus associated with the annulus
and the second being the relative
distance between $u$ and $y$ on one of the boundaries.
The open string state going
through the loop (i.e. annulus) is itself twisted by ${\tilde h}_I$ in the three
planes since the two boundaries
of the annulus sit on two different stacks of D6 branes (say $a=1$ and $a=3$).
We assume here that the
combined system
preserves supersymmetry so that $\sum_I {\tilde h}_I= 0$ mod integers.
Note that as the open string
propagating in the annulus crosses
one of the twist operator insertions it becomes an open string stretched
between the stacks 2 and 3
and when it crosses the second twisted operator it becomes again the string
stretched between 1 and 3.

To compute this correlation function we can go to the torus double cover of the
annulus and take the appropriate square root. The result is
\begin{eqnarray}
D_j F^{(0,2)}_{\bar{i}; \bar{j}}\!\!\!\! &=&\!\!\!\! \int dt^1 dt^2 \int dx \prod_{m=1}^2
\int\!\! d^2z_m\mu_m(z_m)  
\frac{\prod_I \theta_{{\tilde h}_I}[y-Y_I +h_I (u-y)]}{\eta^3\prod_J E(Y_J,u)^{h_J}E
(Y_J,y)^{1-h_J}} \nonumber\\
& & \times\, E(x,y)^{1-\sum_K h_K^2} \;
\epsilon^{bc} \langle\partial X_4(z_b) \partial X_5(z_c) \sigma(u)
\bar{\sigma}(y)\rangle\, ,    
\label{xeq}
\end{eqnarray}
where $Y_3=x$, $Y_4=z_b$ and $Y_5=z_c$. Here, $E(x,y)$ denotes the genus one prime form
\be E(x,y)={\theta_1(x-y)\over\theta_1'(0)}=
{\theta_1(x-y)\over 2\pi\eta^3}\label{eprime}.\ee
Note that in the above equation, $\langle\dots\rangle$
denotes the {\it unnormalized} correlator including
the partition functions of the internal bosons.
The physical string amplitude 
computed by the above quantity is
\begin{equation}
({\cal F}^{+})^2 {\bar\Psi}_{\bar{i}}
{\bar\Psi}_{\bar{j}} \Phi_j\, ,
\end{equation}
where ${\cal F}^+$ is the self-dual field strength and $\Phi$ is the twisted scalar.
In Ref.\cite{agnth} this computation
has been done for the heterotic string and shown to give rise to the topological amplitude above.
The methods of
Ref.\cite{agnth} extend trivially to the open string case, as it can be seen by going to the double
cover of the world-sheet, where the spin structure sum  involves only
one sector (say left-moving sector) exactly as in the heterotic theory.
In the following, we go directly to the broken supersymmetry case
and derive the above topological term in the limit of supersymmetry
restoration, in analogy with the gaugino masses.

We now compute directly the mass term for the fermions
$\Psi$ when supersymmetry is broken
via a VEV of the auxiliary D component of the gauge vector superfield.
Specifically, we take $\sum_I {\tilde h}_I= \epsilon \neq 0$
mod integer, while keeping the supersymmetry condition on $h_I$, namely $\sum_I h_I=1$. In other
words the $h$-twisted sector representing strings stretched between the D6 brane stacks 1 and 2
does not break supersymmetry but the presence of stack 3 breaks it.
This corresponds to the situation when $\Psi$ and its superpartners are massless,
but supersymmetry is broken by the presence of the other
boundary of the annulus associated to the stack 3.
Note that the tree-level mass matrix
does not mix $h$-twisted fermions with the Wilson line fermions.
However we will show below that at one loop
the closed string exchange between the stack 3 and the intersection of 1 and 2 gives
rise to such a mass term.

The amplitude in question is the annulus
three point function of open string states:
\begin{equation}
M_{\bar{i}\bar{j} j}=\langle{\bar\Psi}_{\bar{i}}{{\bar\Psi}_{\bar{j}}} \Phi_j\rangle\, .
\label{mattermass}
\end{equation}
The vertex operators for the fermions in the $(-1/2)$ picture and the scalar
in the $(-1)$ picture are
\begin{eqnarray}
V_j(u) &=& :c e^{-\varphi}\sigma e^{i\sum_I h_I \phi_I}: \nonumber\\
V_{\bar{j}}(y) &=& :c e^{-\frac{\varphi}{2}} e^{\frac{\phi_1-\phi_2}{2}}\bar{\sigma}
e^{i\sum_I (\frac{1}{2}-h_I) \phi_I}: \\
V_{\bar{i}}(x) &=& :e^{-\frac{\varphi}{2}} e^{\frac{-\phi_1+\phi_2}{2}}
e^{i\frac{-\phi_3+\phi_4+\phi_5}{2}}:\, .\nonumber
\end{eqnarray}
We have
inserted the bosonic ghosts $c$ at the vertex $V_j$ and $V_{\bar{j}}$ so we are treating the
surface as twice-punctured
annulus with the associated two moduli (the modulus of the annulus and the relative position
between these two vertices).
The vertex $V_{\bar{i}}$ is dimension 1 and has to be integrated. The total superghost charge of
the three vertices is
$-2$ and therefore we need to insert two picture changing operators $e^\varphi T_F$.
Thus the amplitude
(\ref{mattermass}) becomes
\begin{equation}
M_{\bar{i}\bar{j} j}= \int dt^1 dt^2 \int dx \langle\prod_{m=1}^2
[\int \mu_m (b_L+b_R)] e^\varphi T_F(z_1)
e^\varphi T_F(z_2)V_j(u) V_{\bar{i}}(x) V_{\bar{j}}(y\rangle\, .
\end{equation}

The above amplitude should be independent of the positions $z_1$ and $z_2$ of the
picture changing operators.
Since the total $\phi_4$ and $\phi_5$ charges
of the three vertices is +1 each, it follows that the only relevant
terms in the picture changing operators are
\begin{equation}
e^\varphi T_F\to e^{\varphi} \sum_{I=4,5} e^{-i\phi_I} \partial X_I\, .
\end{equation}
The mass matrix becomes
\begin{eqnarray}
M_{\bar{i}\bar{j} j}&=& \int dt^1 dt^2 \int dx \langle\prod_{m=1}^2
[\int \mu_m (b_L+b_R)]c(u)c(y)\rangle \nonumber\\
&~&\times\sum_s\frac{\theta_s(x-y)^2}{\theta_s(u-\sum z_a + \frac{x+y}{2})}[\prod_I
\theta_{s+{\tilde h}_I}(\frac{x+y}{2} + h_I (u-y) -Y_I)] \nonumber\\
&~&\times~\eta^{-8} (R_{\cal A}\,\makebox{Im}\,\tau)^{-1}\frac{\prod_J E(Y_J,y)^{h_J}E(Y_J,u)^{1-h_J}}
{E(y,u)^{\sum_K h_K^2}E(x,y)E(x,u)E(z_b,z_c)}
\nonumber
\\ &~&
\times ~\epsilon^{bc}\langle\partial X_4(z_b) \partial X_5(z_c) \sigma(u)\bar{\sigma}(y)\rangle \, ,
\end{eqnarray}
where $Y_I$ are as in Eq.(\ref{xeq}). The above formula takes into account
the annulus correction factor $R_{\cal A}$ for the four spacetime bosons, c.f. Eq.(\ref{corgen}), and the
$(\makebox{Im}\tau)^{-1}$ factor due to their zero modes; here, $\tau$ denotes the
usual untwisted modulus of the torus double cover. The power of the $\eta$ function is determined as follows: $\eta^{-4}$ comes from spacetime bosons, $\eta^{-5}$ come from the 5 real scalars that are the bosonization of 10 fermions (spacetime as well as internal), and finally $\eta$ comes from the bosonization of superghost.
In order to perform the spin structure sum over $s$, we choose the
following gauge condition for the positions of the picture changing operators:
\begin{equation}
z_1+z_2-u-y=0\, .
\label{gaugecond}
\end{equation}
With this gauge choice the theta function in the denominator coming from the superghosts cancels
with one
of the theta function coming from the space-time fermions.
After summing over spin structures we obtain:
\begin{eqnarray}
M_{\bar{i}\bar{j} j}&=& \int dt^1 dt^2 \int dx \langle\prod_{m=1}^2
(\int \mu_m (b_L+b_R))c(u)c(y)\rangle \nonumber\\
&~&~~\theta_{\epsilon}(0)[\prod_I
\theta_{{\tilde h}_I-\epsilon}(y-Y_I + h_I (u-y))] \nonumber\\
&~&\times~\eta^{-8} (R_{\cal A}\makebox{Im}\,\tau)^{-1}\frac{\prod_J E(Y_J,y)^{h_J}E(Y_J,u)^{1-h_J}}
{E(y,u)^{\sum_K h_K^2}E(x,y)E(x,u)E(z_b,z_c)}
\nonumber
\\ &~&
\times ~\epsilon^{bc}\langle\partial X_4(z_b) \partial X_5(z_c) \sigma(u)\bar{\sigma}(y)\rangle \, .
\end{eqnarray}
In this equation and in the following $\theta_{\epsilon}$ or $\theta_{{\tilde h}_I-\epsilon}$ denote
the odd theta function twisted by $\epsilon$ or ${\tilde h}_I-\epsilon$.
Note that these twists are purely
along the {\bf a}-cycle ({\bf b}-cycle) in the open (closed) string channel.

In the next step, we
use the bosonization formula for $b,c$ system twisted by $\epsilon$:
\begin{equation}
\langle b(z_1) b(z_2) c(u) c(y)\rangle_{\epsilon} = \frac{\theta_{\epsilon}(z_1+z_2-u-y) E(z_1,z_2)
E(u,y)}{\eta\prod_{m=1}^2 E(z_m,u) E(z_m,y)}.
\end{equation}
Here again, $\langle\dots\rangle$ is the {\it unnormalized} correlator, i.e.\ the complete result of the four-point function in the $b,c$ CFT that also includes its non-zero mode determinant. On the r.h.s., there is $1/\eta$ factor because the bosonization of $b,c$ is one real scalar. Then we can
rewrite the mass term in the form
\begin{eqnarray}
M_{\bar{i}\bar{j} j} &=& \int dt^1 dt^2 \int dx \langle\prod_{m=1}^2
[\int \mu_m (b_L+b_R)]c(u)c(y)\rangle
 H_{\epsilon}(z_1,z_2,u,x,y) \nonumber\\
&& \times ~\frac{\theta_{\epsilon}(0)^2}{\eta^6}(R_{\cal A}\,
\makebox{Im}\,\tau)^{-1},
\end{eqnarray}
where
\begin{equation}
H_{\epsilon}(z_1,z_2,u,x,y) = \frac{\langle \prod_{a=1}^2 G^-(z_a)
V^{\cal T}_{\bar{i}}(x) V^{\cal T}_{\bar{j}}(y)
V^{\cal T}_{j}(u) \rangle^{\cal T}_{\epsilon}}{\langle b(z_1) b(z_2) c(u) c(y)\rangle_{\epsilon}}\, .
\label{He}
\end{equation}
Here we have used the gauge condition (\ref{gaugecond}).
Furthermore in Eq.(\ref{He}), the numerator
is the correlation function in the internal topological theory twisted by $\epsilon$; in particular, this means
that in Eq.(\ref{xeq}), the functions
$\theta_{\tilde{h}_I}$ are replaced by $\theta_{\tilde{h}_I-\epsilon}$.
One can now argue that $H$ does not depend on $z_m$. As a function of $z_1$, both the numerator
and denominator have a first order pole each at $u$ and $y$ and a
first order zero at $z_2$. Each of
them must have one more zero (since the corresponding line
bundle has zero Chern class) but it must
be in the same position as both the line bundles are characterized
by the same twist $\epsilon$ (i.e.
the same point in the Jacobian torus). Since $H$ as a function of $z_1$ is a section of the
trivial line bundle and has no zero or pole, it must be constant. A similar argument applies for
$z_2$. Therefore
\begin{equation}
H_{\epsilon}(z_1,z_2,u,x,y) = B_{\epsilon}(u,x,y)\, .
\end{equation}

Now let us take the $\epsilon \rightarrow 0$ limit
and compute the leading term in $\epsilon$. Since in the closed string channel
the twist $\epsilon$ is purely along the {\bf b}-cycle, in this limit  $\theta_{\epsilon}(0) \rightarrow
\epsilon \eta^3$. The correction factor for spacetime bosons with Neumann boundary conditions yields $(R_{\cal A}\,
\makebox{Im}\,\tau)^{-1}=1$, see  Eq.(\ref{corfin}).
The leading contribution is therefore of order $\epsilon^2$, and at
this order we can set $\epsilon=0$ in $B_{\epsilon}(u,x,y)$. Finally $B_{\epsilon=0}$
multiplied by the
ghost correlators involving the Beltrami differentials effectively replaces $\int \mu b$ by
$\int \mu G^-$ which gives the topological quantity $D_j F^{(0,2)}_{\bar{i}; \bar{j}}$.
The final result, to order $\epsilon^2$,  therefore is
\begin{equation}
M_{\bar{i}\bar{j} j} = \epsilon^2  D_j F^{(0,2)}_{\bar{i}; \bar{j}}\, .
\label{masstop}
\end{equation}

Eq.(\ref{masstop}) determines the Yukawa type coupling (\ref{mattermass}) involving two
anti-chiral fermions and one chiral boson. Note that this coupling cannot be derived from
a superpotential and is not  allowed by supersymmetry. Here, it was induced by
a supersymmetry breaking D-term. If the twisted scalar $\Phi_j$ acquires a VEV, it
generates a Dirac mass term mixing the bi-fundamental fermions 
with a Wilson line fermion. In fact, such a VEV breaks also the gauge group,
generating further mass mixing of the bi-fundamental fermions with gauginos
through gauge Yukawa couplings.
Since the right hand side term of Eq.(\ref{masstop})
appears in the holomorphic anomaly of $F^{(0,3)}$, which in turn
gives the gaugino mass at this order, we conclude that the corresponding
fermion mass matrix elements are given
by the holomorphic anomaly of the gaugino mass term.

One can further ask what happens when one takes anti-holomorphic derivative with respect to
some moduli say $\partial_{\bar{i_2}}$ of
$F^{(0,2)}_{\bar{i_1}; \bar{j_1}}$. From the analysis of section 4, after anti-symmetrizing
in ${\bar i}_1$ and ${\bar i}_2$, we find that the result again comes
from various degeneration limits. In particular, the degeneration corresponding to
open-string intermediate states gives rise to $F^{(0,1)}_{\bar{i_1}\bar{i_2};\bar{j_1}\bar{j_2}}$
associated to a $\Pi^2$ term in the effective action at the disk level.
For instance the indices ${\bar i}_1$ and ${\bar i}_2$ can refer to
some open string moduli fields, while ${\bar j}_1$ and ${\bar j}_2$
can refer to bi-fundamental open string states.
This coupling can be evaluated by a 6-point function on a disk,
involving two pairs of twist-antitwist fields
and the two moduli fields, corresponding to $D_{j_1} D_{j_2}
F^{(0,1)}_{\bar{i_1}\bar{i_2};\bar{j_1}\bar{j_2}}$.

\section{A Simple Example}

In this section, we present a simple toroidal example in which the topological partition function
$F^{(0,3)}$, as well as all lower order quantities appearing in the holomorphic anomaly
equations, can be computed either explicitly or by using some symmetry arguments.

Our starting point is a configuration in which every two brane stacks of the three
boundaries intersect nontrivially only in two out of the three internal planes and are
parallel in the remaining one, a configuration different from the one considered
in the previous section. Furthermore, to avoid the enhancement of supersymmetry
to $N=2$, and thus the vanishing of $F^{(0,3)}$, the plane in which the branes are
parallel must be different in every of the three possible pairs. First we choose the
horizontal axis in each plane along the stack $a=3$. Then, we pick stacks
2 and 3 to be parallel in the plane $I=3$, stacks 1 and 3 to be parallel in the plane
$I=4$ and  stacks 1 and 2 to be parallel in the plane $I=5$. This is described by
the brane angles:
\be
\theta^3_I=0\qquad \theta^1_4=\theta^2_3=0\quad \theta^1_5=\theta^2_5\qquad
\theta^1_3+\theta^1_5=0\quad \theta^2_4+\theta^2_5=0\, ,
\label{anglesex}
\ee
where the last two relations follow from space-time supersymmetry. According to
Eq.(\ref{twistangle}), the corresponding orbifold twists along the two {\bf b}-cycles are:
\be
(g_3^1,g_3^2)=(e^{4\pi i\theta},1)\quad (g_4^1,g_4^2)=(1,e^{-4\pi i\theta})\quad
(g_5^1,g_5^2)=(e^{-4\pi i\theta},e^{4\pi i\theta})\, ,
\label{twistsex}
\ee
where the angle $\theta=\theta^1_3$ is an arbitrary parameter.

Now the constraint  (\ref{npqconst}) on the lattice sum is solved trivially,
leading for each plane to a summation over two unrestricted integers of the corresponding
two-dimensional momentum lattice depending on the complex structure modulus $U_I$
and Wilson line $A_I$. Indeed, when two branes are parallel within a plane $I$,
say the stacks $a=1$ and $a=2$, the corresponding magnetic fluxes $p_a^I/q_a^I$ are
equal and since $(p_a^I,q_a^I)$ are relatively prime, one has $p_1^I=p_2^I$ and
$q_1^I=q_2^I$. Eq.(\ref{npqconst}) then requires
$p_1^I({\vec n}_I+{\vec n}'_I)+p_ 3^I{\vec n}''_I=0$ and
$q_1^I({\vec n}_I+{\vec n}'_I)+q_3^I{\vec n}''_I=0$, implying
${\vec n}_I+{\vec n}'_I={\vec n}''_I=0$ since the third stack must have non-trivial
intersection with the other two. One is then left with a summation over two unrestricted
integers, defined for instance by the vector ${\vec n}_I$. Note that the the physical Wilson line
${\vec\alpha}$
is given by the difference
${\vec\alpha}^I={\vec\alpha}^I_1-{\vec\alpha}^I_2$ and corresponds in the T-dual picture to the
relative distance between the two parallel brane stacks.

Finally, there is an $SL(2,\mathbb{Z})_I$ action on each $U_I$ and $A_I$:
$U_I \to(aU_I+b)/(cU_I+d)$ and $A_I \to A_I/(cU_I+d)$. The modular weights of
$F^{(0,h)}_n$ are determined by their K\"ahler weights $n+h-2$ \cite{agnth}. Thus
$F^{(0,3)}$ transforms with weight 1 under each $SL(2,\mathbb{Z})_I$ and is also
monodromy invariant under $A_I \to A_I+1$ and $A_I \to A_I+U_I$. Furthermore,
it vanishes when $A_I =1/2, U_I/2, (U_I+1)/2$. By taking derivatives
with respect to $A_I$ or $\bar{A}_I$ and setting $A_I=0$ one finds
that the zeroes are of first order. On the other hand, for $A_I =0$,
the two stacks become coincident and there are additional massless states.
As a result, $F^{(0,3)}$ is singular at the origin and acquires a first order pole
instead of a zero as in the other points. A simple ansatz for $F^{(0,3)}$, satisfying
all properties above, is:
\be
F^{(0,3)}=f_3\prod_I H(U_I,A_I)\quad;\quad H(U,A)={\theta'(A)\over\theta(A)}+2i\pi
{{\rm Im}A\over{\rm Im}U}\, ,
\label{ansatz}
\ee
where $f_3$ is a numerical constant and the prime denotes differentiation
with respect to $A$. Actually, this expression, which will be verified subsequently
by studying the holomorphic anomalies, also suggests that for this special
brane configuration described by the orbifold twists (\ref{twistsex}), the integration
domain of the three twisted world-sheet moduli $\tau_{\{g_I\}}$ is factorized into
three independent integrals over the positive real line, so that each one can be
performed explicitly yielding the result (\ref{ansatz}):
\be
\int_0^\infty dl f(U,{\vec a},l)={1\over\pi}{\sum_{n_1,n_2}}^\prime
{e^{2i\pi{\vec n}{\vec a}}\over n_1+n_2U}=-{1\over\pi}H(U,A)\, ,
\label{intf}
\ee
where the function $f$ is given in Eq.(\ref{F03f}). An appropriate $SL(2,\mathbb{Z})$ invariant
regularization of the above sum leads to the r.h.s. part of the equation, with $H$
acquiring a non-holomorphic dependence as in (\ref{ansatz}).

Our strategy to prove Eq.(\ref{ansatz}) will be to compute the holomorphic ano\-maly
as discussed in section 4 and show that the latter is reproduced by (\ref{ansatz}).
Holomorphic ambiguity is then fixed by requirement of target space $SL(2,\mathbb{Z})_I$
and $A_I$ monodromy properties. Taking a derivative of (\ref{ansatz}) with respect to an
anti-holomorphic open string Wilson line,
for instance ${\bar A}_3$, one finds
\be
\label{32holanex}
\partial_{{\bar A}_3}F^{(0,3)} = -{\pi f_3\over{\rm Im}U_3}\prod_{I=4,5}H(U_I,A_I)\, .
\ee
On the other hand, from the general discussion of section 4 on holomorphic anomaly,
the contribution comes from various degeneration limits of the surface. To analyze this we
need to study the behavior of the twisted $\tau_{\{g_I\}}$ in the three degeneration limits
as shown in Fig.~4. The case that we are considering is characterized by the twists
(\ref{twistsex}). It is more convenient to normalize the three twisted differentials
$\omega_{\{g_I\}}$ along the periods shown in Fig.~6 of the Appendix, so that
$\int_{{\bf a}_2}\omega_{\{g_3\}} = \int_{{\bf a}_1}\omega_{\{g_4\}} =
-\int_{{\bf a}_1{\bf a}_2}\omega_{\{g_5\}} = 1$. The twisted $\tau_{\{g_I\}}$ are then defined as
$\tau_{\{g_3\}}=\int_{{\bf b}_2}\omega_{\{g_3\}}$,  $\tau_{\{g_4\}}=\int_{{\bf b}_1}\omega_{\{g_4\}}$
and $\tau_{\{g_5\}}=\int_{{\bf b}_1{\bf b}_2}\omega_{\{g_5\}}$. Note that for the twists (\ref{twistsex}),
${\bf b}_2$, ${\bf b}_1$ and ${\bf b}_1{\bf b}_2$ are closed cycles
for $\omega_{\{g_I\}}$ for $I=3,4,5$ respectively.

Now let us consider the three degeneration limits shown in Fig.~4.
\begin{enumerate}
\item In the first one where
the genus 2 double cover degenerates along the dividing geodesic corresponding to an
open string intermediate state between two annuli whose double covers have
cycles $({\bf a}_1,{\bf b}_1)$ and $({\bf a}_2,{\bf b}_2)$,
$\omega_{\{g_4\}}$ and $\omega_{\{g_3\}}$ degenerate to the untwisted
differentials of the two torii respectively while $\omega_{\{g_5\}}$ degenerates to twisted
differentials on the two torii having first order pole at the node.
Thus while $\tau_{\{g_4\}}$ and
$\tau_{\{g_3\}}$ are finite (they are just the untwisted moduli associated with the two
torii), $\tau_{\{g_5\}}$ becomes infinite. In fact in terms of the plumbing fixture
coordinate, say $t$, $\tau_{\{g_5\}}$ goes as $-i\ln|t|$. The two other such degenerations are
obtained by permutations where either $\tau_{\{g_3\}}$ or $\tau_{\{g_4\}}$ goes to infinity
keeping the remaining two finite. Taking derivative with respect to ${\bar A}_3$ and using
Eq.(\ref{ident}), we note that the relevant such degeneration limit comes from
$\tau_{\{g_3\}}$ going to infinity. The resulting contribution to the holomorphic anomaly is
\be
\partial_{{\bar A}_3}F^{(0,3)}=
\sum_{I=4,5}F^{(0,2)}_{{\bar A}_3;{\bar A}_I}G^{{\bar A}_IA_I}\partial_{A_I}F^{(0,2)}\, .
\label{32holan}
\ee

\item In the second degeneration limit which results in a twice punctured annulus with
an open string intermediate state, one of the $\tau_{\{g_I\}}$ goes to zero keeping the
remaining two finite. For instance if in Figure 6,  we move the ${\bf a}_2$ cycle near the
real axis (i.e. the third boundary), then ${\bf b}_2$ cycle shrinks to zero so that $\tau_{\{g_3\}}$
goes to zero, but $\tau_{\{g_4\}}$ and $\tau_{\{g_5\}}$ remain finite. $\tau_{\{g_4\}}$
becomes the usual untwisted modulus $\tau$ associated with the resulting
annulus while $\tau_{\{g_5\}}$ is a function of $\tau$ and the separation between the two
punctures. In our case, however, this degeneration limit does not contribute so long as
every pair of brane stacks is separated (by suitable Wilson line) in the plane in which they
are parallel to each other. This gives masses to intermediate open strings that are stretched
between different pairs of stacks and hence this degeneration limit is exponentially
suppressed.

\item In the third degeneration limit with intermediate closed string states is obtained
by shrinking one of the ${\bf a}$ cycles in Fig.~6. For instance if we shrink ${\bf a}_2$ cycle
to a point $P$ $\tau_{\{g_3\}}$ clearly goes to infinity
(going as $-i\ln|t|$ with $t$ being the corresponding
plumbing fixture coordinate).  In this limit $\omega_{\{g_4\}}$ becomes untwisted
differential on the remaining torus (double cover of the annulus with boundaries
${\bf a}_1$ and ${\bf a}_3$) and hence becomes
the usual untwisted modulus of the annulus in the closed string
channel. $\omega_{\{g_5\}}$ on the other hand remains a twisted differential on the torus
with single poles at the points $P$ and its image $P'$. As a result  $\tau_{\{g_5\}}$
goes to infinity as $-i\ln|t|$. This degeneration limit does not contribute to the
holomorphic anomaly due to the fact that two of the $\tau_{\{g_I\}}$ go to infinity
simultaneously in this limit. This is because the integrand in (\ref{F03}) appears with
one momentum each from each plane through $f$ defined in (\ref{F03f}). Explicitely this factor
is $\prod_I (n_1^I+n_2^I \bar{U}_I)$. The momentum in
one of the planes ($I=3$) disappears when one takes the derivative with respect to
${\bar A}_3$ due to the identity (\ref{ident}), however the other two momentum factors for
$I=4,5$ still exist in the integrand. Since in this closed string degeneration limit
$\tau_{\{g_3\}}$ as well as another one of the $\tau_{\{g_I\}}$ for $I=4$ or $I=5$ go
to infinity, we conclude that this limit is exponentially suppressed.
\end{enumerate}
Thus the holomorphic anomaly is entirely given by the first degeneration limit
(\ref{32holan}).  $F^{(0,2)}_{{\bar A}_3;{\bar A}_I}$ appearing in this equation comes
from the effective action term
$\Pi{\rm Tr}W^2$ and we will show in the following that it is given by:
\be
F^{(0,2)}_{{\bar A}_I;{\bar A}_J}={f_2^{IJ}\over{\rm Im}U_I{\rm Im}U_J}H(U_K,A_K)
\quad ;\quad I\ne J\ne K\ne I\, ,
\label{F02piex}
\ee
where $f_2^{IJ}$ are numerical constants.

The function $F^{(0,2)}$ appearing in Eq.(\ref{32holan}) is the one loop
gauge kinetic function providing the threshold corrections to gauge
couplings ${\rm Tr}W^2$~\cite{apt} .
Its Wilson line dependence receives contributions only from $N=2$ supersymmetric sectors
and reads:
\be
\partial_{A_I}F^{(0,2)}=b_I\partial_{A_I}\Delta(U_I,A_I)\quad;\quad
\Delta(U,A)=\ln|\theta_1(A)|^2-2\pi{({\rm Im}A)^2\over{\rm Im}U}\, ,
\label{gcex}
\ee
where $b_I$ are numerical coefficients related to the $N=2$ beta-functions and
$\partial_A \Delta=H$. Using (\ref{F02piex}) and the Wilson line metric
$G_{A_I{\bar A}_I}=1/{\rm Im}U_I$, one can then identify (\ref{32holan})
with (\ref{32holanex}), implying $-\pi f_3=\sum_{I=4,5}f_2^{3I}b_I$.
Note that for our choice of angles (\ref{anglesex})-(\ref{twistsex}), placing
at the two boundaries of the annulus two different brane stacks one obtains
an $N=2$ sector associated to the plane where the two stacks are parallel.

We now compute $F^{(0,2)}_{{\bar A}_I;{\bar A}_J}$ from the four-point annulus
amplitude involving two gauge fields and two anti-chiral fermions ${\bar\chi}_I$,
${\bar\chi}_J$, that belong to the corresponding Wilson line supermultiplets.
The gauge boson vertices are given in Eq.(\ref{vg}), while the ${\bar\chi}_I$
fermion vertex at zero momentum, in the $-1/2$ ghost picture, is given by:
\be
V^{(-1/2)}_{{\bar\chi}_I^\alpha}=:e^{-\varphi/2}S_\alpha S_I:\quad;\quad
S_I=e^{{i\over 2}(\phi_I-\phi_J-\phi_K)}\quad I\ne J\ne K\ne I\, ,
\label{Vchi}
\ee
where we use the notation of section 3. Choosing as fermion vertices, say
$I=3$ and $J=4$, one should also insert a picture changing operator $e^\varphi T_F$,
from which only the term $e^\varphi e^{i\phi_5}\partial{\bar X}^5$,
from the supercurrent (\ref{tfint}),
contributes. Evaluating the correlator at the quadratic order in external momenta
and performing the spin-structure sum is straightforward. All non-zero mode
determinants cancel and one is left over with a summation over the lattice
momenta of the torus $I=5$, associated to the $N=2$ sector defined by
the first two brane stacks. The result is:
\be
F^{(0,2)}_{{\bar A}_I;{\bar A}_J}={f_2^{IJ}\over{\rm Im}U_I{\rm Im}U_J}
\int_0^\infty dl f(U_K,{\vec a}_K;l)\, ,
\label{F02pires}
\ee
where the function $f$ is defined in Eq.(\ref{F03f}). The integral can be performed
as in (\ref{intf}), leading to (\ref{F02piex}). These results prove our
initial ansatz (\ref{ansatz}). In fact, the same expression can be obtained by
integrating the holomorphic anomaly equation (\ref{32holan}) by using the explicit
forms for $F^{(0,2)}_{{\bar A}_I;{\bar A}_J}$ and $F^{(0,2)}$ and $SL(2,\mathbb{Z})$
symmetry.

Now we turn to the holomorphic anomaly of $F^{(0,2)}_{{\bar A}_I;{\bar A}_J}$.
Taking an anti-holomorphic derivative of (\ref{F02piex}) with respect to ${\bar A}_K$,
one finds:
\be
\partial_{{\bar A}_K}F^{(0,2)}_{{\bar A}_I;{\bar A}_J}=-
{\pi f_2^{IJ}\over{\rm Im}U_I{\rm Im}U_J{\rm Im}U_K}\, .
\label{dbarF02}
\ee
On the other hand, using the integral form (\ref{F02pires}) and the identities (\ref{ident}),
one obtains a contribution only from the origin of the lattice at the boundary $l=\infty$.
Since this corresponds to the infrared limit in the closed string channel, one
concludes that in this case it is the closed string degeneration limit that contributes
and the holomorphic anomaly equation becomes:
\be
\partial_{[{\bar A}_K}F^{(0,2)}_{{\bar A}_I;{\bar A}_J]}=
F^{(0,1)}_{{\bar A}_K{\bar A}_I;{\bar A}_J{\bar\Phi}}D_\Phi F^{(0,1)}\, ,
\label{dbarF02an}
\ee
where $\Phi$ is a closed string modulus. Actually, $F^{(0,1)}$ must be the (tree)
gauge kinetic function on the disk, implying that $\Phi$ is the string dilaton which
couples on the disk but not on the torus. The first factor in the r.h.s. of (\ref{dbarF02an})
must then correspond to a $\Pi^2$ term on the disk. It can be computed independently
by a four-point amplitude involving two anti-chiral fermions, say $I$ and $J$, and two
scalars, the Wilson line ${\bar A}_K$ and the dilaton, at a level quadratic in the external
momenta.

Let us close the section by giving a possible form of $F^{(0,3)}$ in the case
of non-parallel brane stacks. In this case, $F^{(0,3)}$ is monodromy invariant under
$A_I \to A_I+1/p_I$ and $A_I \to A_I+U_I/p_I$, where $p_I$ is defined in Eq.(\ref{pi}).
Furthermore it vanishes when any of the $p_I A_I = 0, 1/2, U_I/2, (U_I+1)/2$. Unlike
the previous case, now there is no pole at the origin because at this point no new massless
states emerge. As before, by taking derivatives
with respect to $A_I$ or $\bar{A}_I$ and setting for instance $A_I=0$,
one finds that the zeroes are of first order.
We assume the factorized form:
\be
F^{(0,3)}=\prod_I \partial_{A_I} G\, ,
\ee
where $G$ is real modular invariant function. A possible ansatz is
\be
G = \prod_I G_I~; ~ G_I \!=\! ({\rm Im} U_I) e^{-2\pi \frac{{\rm Im} p_I^2 A_I^2}{{\rm Im}U_I}}
[c_o|\theta_1(p_I A_I;U_I)|^2 + c_e \sum_e |\theta_e(p_I A_I;U_I)|^2 ]
\ee
where subscripts $e$ and $o$ refer to even and odd spin structures respectively and $c_o$ and
$c_e$ are real $SL(2,\mathbb{Z})_I$ invariant functions of $U_I$ which may also depend on the
brane angles. We could try to check this ansatz by computing the holomorphic
anomaly of $F^{(0,3)}$
as before. Taking derivative with respect to $\bar{A}_3$ one again gets contribution
only from the degeneration limits. This time however, in the first as well as the third
degeneration limits all the three twisted $\tau_{\{g_I\}}$ go to infinity due to
the fact that the resulting torii are twisted in all planes. These limits therefore
are exponentially suppressed. The only contribution comes from the second degeneration
limit where the resulting twice punctured annulus comes with the insertion of twisted
states at the punctures. This is the quantity we considered in section 7. However
an explicit computation of this term, although in principle possible,
is considerably more difficult and we shall not
pursue it here.

\section{Concluding Remarks}

To summarize, in this work we discussed the identification of the topological
partition function $F^{(0,h)}$ of
the two-dimensional $N=2$ twisted Calabi-Yau $\sigma$-model (in the orbifold limit),
on bordered  genus-zero world-sheets with $h$ boundaries,
with the moduli-dependent couplings associated to the F-terms $({\rm Tr}W^2)^{h-1}$,
where $W$ is the familiar gauge superfield appearing in the effective four-dimensional
$N=1$ supersymmetric effective action of the Type I string theory compactified on
the same six-dimensional orbifold.

We then studied the holomorphic anomaly equation for the violation of
the expected holomorphicity of $F^{(0,h)}$. The anomaly equation involves also the correlation functions $F_n^{(0,h-1)}$ associated
to F-terms of the form $\Pi^n({\rm Tr}W^2)^{h-2}$, where $\Pi$ denotes a generic
chiral projection of a non-holomor\-phic functions of chiral superfields. This system is
similar to the heterotic string case studied in the past and related to it by S-duality. A remaining open problem is the integrability of this equation, which
can be shown, as in the heterotic string, only in the absence of the handle
degeneration limit. Its non-integrability in the general case suggests that there may
be still more ``topological" quantities missing, that are required for the closure of
the full topological theory on world-sheets with boundaries.

Another interesting open problem is the generalization of the possible physical
interpretation in terms of actual string amplitudes of the topological partition
function $F^{(g,h)}$ for higher genus $g>0$.
A naive extension of our results, following the methods presented in this work,
to world-sheets with handles and boundaries simultaneously, fails because the
zero-mode contributions of space-time coordinates do not cancel, and thus, the
measure does not become topological.

An important phenomenological application of our results is in theories
with supersymmetry broken by the VEVs of auxiliary D-components of gauge vector supermultiplets. Such a breaking appears, in particular, in the presence of non-trivial
internal magnetic fields, or equivalently in the T-dual description, of branes
intersecting at angles, and can be studied directly at the string level.
Then the effective operators $({\rm Tr}W^2)^2$
and $\Pi {\rm Tr}W^2$ generate fermion masses and the associated
topological functions become physically observable in the mass spectrum. The resulting Majorana gaugino masses are given by
$F^{(0,3)}$ at two loops, while the matter fermion masses are of Dirac type and are given
by the topological quantity $F^{(0,2)}_{{\bar i};{\bar j}}$, appearing at one loop
in the holomorphic anomaly equation of $F^{(0,3)}$.
They are both of order
$m_0^4/M_S^3$ for $m_0<M_S$, where $m_0$ is the supersymmetry breaking
scale and $M_S$ is the string mass.  We presented simple examples of toroidal
string compactifications with intersecting branes, where both quantities can be
computed explicitly as functions of the closed string geometric moduli and of
open string Wilson lines.

The precise implementation of these results in a complete string framework,
including the Standard Model and a consistent mechanism of supersymmetry
breaking and moduli stabilization, remains of course an open issue.
For instance, an important question is the effect of closed string back-reaction
on the supersymmetry breaking induced in the open string sector by turning on
internal magnetic fields, or equivalently by the presence of D-branes at angles.
A possible application is in the recently proposed scenario of split
supersymmetry~\cite{split}.
Our results then provide a natural mechanism to break R-symmetry
by string effects and generate gaugino and higgsino masses at the TeV scale,
when squark and slepton masses are at high energies, of the order of
$10^{13}$ GeV, and $M_S$ is near the unification scale of $10^{16}$ GeV.
Alternatively, our results can be used to generate gaugino and non-chiral
fermion masses in the context of low string mass scale and large extra
dimensions~\cite{low}, when supersymmetry is broken at the string scale
by appropriate configurations of branes and orientifolds~\cite{bsb}.
\section*{Acknowledgments}
We would like to thank M.S.Narasimhan for valuable discussions on Prym differentials.
This work was supported in
part by the European Commission under the RTN contracts
MRTN-CT-2004-503369 and MEXT-CT-2003-509661, in part by the
INTAS contract 03-51-6346, and in part by the CNRS PICS \# 2530.
The research of T.R.T.\ is supported in part by the U.S.\ National Science
Foundation Grant PHY-0242834.
K.S.N. and T.R.T. thank the CERN theory division and I.A. thanks ICTP for hospitality,
during multiple visits.
T.R.T.\ is grateful to Dieter L\"{u}st, Stephan Stieberger and
to Arnold Sommerfeld Center for Theoretical Physics at Ludwig Maximilians
University in Munich for their kind hospitality.
Any opinions, findings, and conclusions or recommendations expressed in this material are those of the authors and do not necessarily reflect the views of the National Science Foundation.

\appendix
\section*{Appendix}
\section{Lattice Contribution to the Amplitudes}

In this Appendix, we derive the lattice sums involved in the topological partition function
$F^{(0,h)}$. Since we are considering product of
three planes (tori) with fluxes, it is sufficient to focus on one plane. In order to compute the
lattice contribution, it is more convenient to go to the T-dual theory where the D9 branes become
D6 branes whose world-volumes span lines in each of the 3 internal planes. These lines are parallel to some lattice vectors. They
are at angles given by the fluxes in the corresponding plane.
Let $\vec{v}_i$ be a primitive lattice vector parallel to the $i$-th brane and let $\vec{w}_i$
be such that ($\vec{v}_i, \vec{w}_i)$, for each $i$, spans the two-dimensional lattice describing
the torus. This in particular means that $v_i.*w_i \equiv v_i^{\mu} w_i^{\nu} \epsilon_{\mu\nu}=\pm T_2$ where
$T_2=\sqrt{G}$ is the area of the basic cell of the lattice (in the following we choose
the orientation of $w_i$
so that the sign in this equation is plus). We can use a complex coordinate $Z$ to describe the
plane. Let $v_i$ and $w_i$ be the complex numbers representing
the lattice vectors $\vec{v}_i$ and $\vec{w}_i$, respectively. We are considering a
world-sheet of genus zero and $h$ boundaries.
For this bordered surface, we define the ${\bf a}_i$ cycles,
$i=1,2,\dots,h$, as the boundaries $\alpha_i$ shown in Fig. 1 (of course, these
cycles are not independent: they satisfy $\prod_{i=1}^h {\bf a}_i =\bf{1}$).\footnote{These cycles should not be confused with the basis of $\bf a$-cycles used in Section 2 and depicted in Fig. 2.}
Let $P_i$ be a point on the ${\bf a}_i$ cycle. Since  ${\bf a}_i$ lies on the $i$-th brane, $Z(P_i)$
takes values
\be
Z(P_i) = x_i(P_i) v_i + (m_i +y_i)w_i
\label{bc}
\ee
where $x_i$ and $m_i$ are arbitrary real and
integer numbers, respectively, and $y_i$ is a {\it fixed} real number such that
$|y_i|\leq 1/2$ which describes
the relative transverse position of the $i$-th brane.\footnote{In the
D9 brane theory $y_i$ is one of
the components of the Wilson line which after T-duality represents the transverse position of the
brane. The second component of the Wilson line  remains a Wilson line on the D6 brane.}
We are considering the case where not
all the branes are parallel. Let us assume that the $(h-1)$-th and the
$h$-th branes are not parallel to
each other. We can choose a
complex coordinate $Z$ for the plane such that the intersection of these two branes is at
$Z=0$ and furthermore we can rotate the coordinate so that the $h$-th
brane lies on the real axis. With this
choice of coordinate, we have set $v_h$ a real number and
$m_{h-1}=m_h=y_{h-1}=y_h=0$. The remaining
integers $m_i$ for $i=1,\dots,h-2$ are arbitrary and result in a ($h-2$)-dimensional lattice sum.
However, one can easily see
using the fact that $(v_i, w_i)$ span the lattice for each $i$, that this change of
coordinates results in the shift
\be
Z(P_i) = x_i(P_i) v_i + (m_i + {\tilde{y}}_i - \frac{v_h.*v_i}{v_h.*v_{h-1}}m_{h-1})
w_i~~; ~~ i=1,\dots,h-2
\ee
where ${\tilde{y}}_i$ are the effective
Wilson lines in the Dirichlet direction
(or transverse positions in the T dual picture) and are given by
\be
{\tilde{y}}_i = y_i - y_{h-1} \frac{v_h.*v_i}{v_h.*v_{h-1}} - y_h
\frac{v_{h-1}.*v_i}{v_{h-1}.*v_h}\, .
\label{wilsond}
\ee
Note that the ratio $\frac{v_h.*v_i}{v_h.*v_{h-1}}$ is a rational number.
Let $p$ be the smallest integer such that $p$ times this rational number,
for each $i$, is integer. Then
$m_{h-1}=0,1,\dots,p-1$ give the different conjugacy classes of the lattice sum over $m_i$.

As one goes with points $P_i$
around the cycles ${\bf a}_i$, the functions $x_i(P_i)$ can shift by integers: $x_i \rightarrow x_i + n_i$. The
integers
$n_i$ denote the winding of the string on the boundaries. Not all the
integers $n_i$ are independent,
however. This is because $\prod_{i=1}^h {\bf a}_i$ is homotopically trivial, giving rise to the condition
\be
\sum_{i=1}^h n_i v_i =0.
\label{nicons}
\ee
Since this is a complex equation with not all the $v_i$ parallel,
there are only $h-2$ independent
integers, say $n_i$ for $i=1,\dots,h-2$.
It is important to note that $n_i$ actually span only a sublattice of integers
such that there exist $n_{h-1}$ and $n_h$ satisfying Eq.(\ref{nicons}).
Since $(v_h,w_h)$ generate the
lattice, $v_i = p_i v_h + q_i w_h$, $i=1,\dots,h-1$, for some coprime
integers $(p_i, q_i)$. Therefore choosing $n_h = -\sum_{i=1}^{h-1} n_i p_i$
the above constraint on $n_i$ becomes equivalent to the constraint
\be
\sum_{i=1}^{h-1} n_i (v_h.*v_i) \equiv \sum_{i=1}^{h-1} (1-g_i^{-1})n_i v_i = 0\, ,
\ee
where $g_i = e^{2i \theta_i}$ with $\theta_i = arg(v_i)$ and we have used the fact that, for our
choice of coordinate $Z$, $v_h$ is real.

The boundary conditions on the branes can be conveniently imposed by going to the double cover of
$\Sigma_{(0,h)}$ which is a genus $g=h-1$ Riemann surface, with an
anti-analytic $\mathbb{Z}_2$ involution that
keeps the ${\bf a}_i$ cycles fixed and takes ${\bf b}_i$ cycles to ${\bf b}_i^{-1}$, see Section 2.
Then going around ${\bf b}_i$ cycle $dZ \rightarrow g_i dZ$, while around
${\bf a}_i$ cycles $dZ$ is invariant.
$\mathbb{Z}_2$ involution which takes a point $P$ to $P'$ acts on $Z$ as
follows. Let $P_0$ be a base point on the ${\bf a}_h$ cycle lying on the $h$-th brane. By our choice of
coordinate, $Z(P_0)$ is real. We can define
$Z(P) = Z(P_0)+\int_{P_0}^P dZ$. $Z(P)$ is not single-valued -- it
depends on the homotopy class $H(P_0,P)$
of the path chosen. $\mathbb{Z}_2$ involution acts as $Z(P') = \bar{Z}(P)$
where $Z(P')$ is defined through a
path that is in the homotopy class $G H(P_0, P')$, with $G$ being
the $\mathbb{Z}_2$ action on the homotopy class.
Clearly, $Z$ on the ${\bf a}_h$ cycle is real. Now consider a
point $P_i$ on the ${\bf a}_i$ cycle. Then $Z(P_i)$ is given by
\begin{equation}
Z(P_i)-Z(P_0) = \int_{P_0}^{P_i} dZ = \int_{b_i} dZ + g_i \int_{P_0}^{P_i}
d\bar{Z} = \int_{b_i} dZ + g_i (\bar{Z}(P_i)-Z(P_0)).
\end{equation}
The general solution of this equation is precisely of the form (\ref{bc}).

Now we would like to write the classical solutions for $Z$ for a
given set of integers $m_i$, $n_i$
and the transverse positions $y$. On genus $h-1$ surface there are $h-2$ linearly independent
holomorphic twisted (Prym)
differentials $\omega_{j,\{g\}}$ for $j=1,...,h-2$, that are twisted
by $g_i$ around ${\bf b}_i$ cycles and untwisted
around ${\bf a}_i$; here, the subscript $\{g\}$ denotes the collection of
twists $\{g_i\}$. Let us choose a marking
for the surface as shown in Fig. 6.\footnote{Fig. 6 is just the double cover of Fig. 1
and the
corresponding contractible surface obtained by cutting along the lines shown in Fig. 6
is just the double cover
of Fig. 3. In the following, for notational simplicity,
we will choose the ${\bf a}$-cycles to be the $h-1$ of the boundaries.
This is different from the convention followed in Section 2
but it has the advantage of directly giving the lattice momenta of the boundary states.}
\begin{figure}[h]\hfill
\includegraphics[scale=.6]{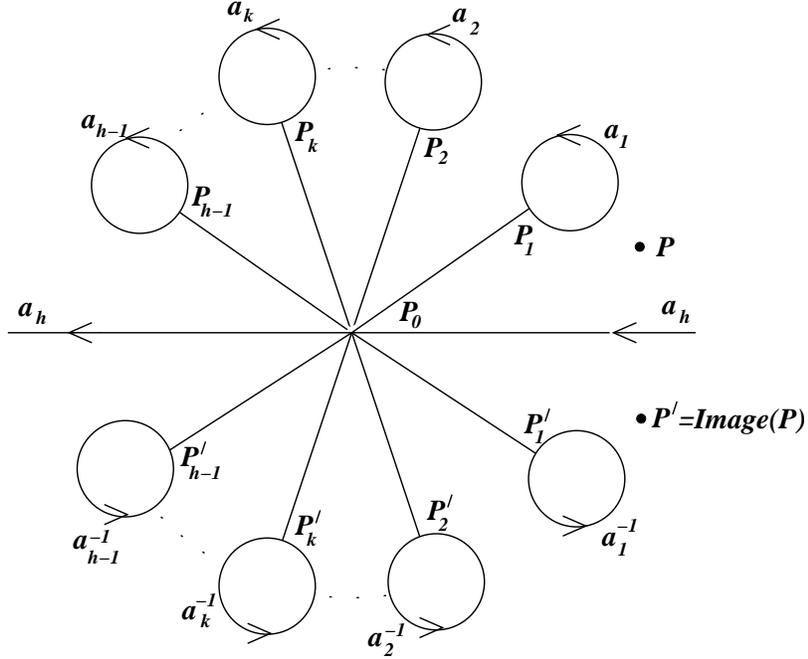}\hfill\hfill
\caption{$\Sigma_{(0,h)}$ is represented in the upper half-plane with $h$
boundaries, $\bf{a}_h$ being the real axis and the remaining $\bf{a}_1,\dots,\bf{a}_{h-1}$
being the boundaries of $h{-}1$ holes cut out. The double cover is obtained
by including the lower half-plane with the $\mathbb{Z}_2$ involution
$z\to\bar z$. The images of boundaries $\bf{a}_1,\dots,\bf{a}_{h-1}$ are shown as
$\bf{a}_1^{-1},\dots,\bf{a}_{h-1}^{-1}$ while $\bf{a}_h$ is fixed. $\bf{a}_k$ and $\bf{a}^{-1}_k$,
for each $k=1,\dots,h{-}1$, are glued together to yield a genus $h{-}1$
surface. Simply-connected surface is obtained by cutting lines $P_0$ to
$P_k$ and $P_0$ to $P_k'$ for each $k$. ${\bf b}_k$ cycle is defined as the path
from $P_0$ to $P_k$ which is identified with $P_k'$, and then from $P_k'$ to $P_0$,  (in a different sheet due to the twist along $\bf b_k$ cycles).}
\label{topf6}
\end{figure}
One gets the genus $h-1$ surface by gluing ${\bf a}_i$ and
${\bf a}_i^{-1}$ cycles together. The real axis is the boundary
${\bf a}_h$ which sits on the $h$-th brane while
the remaining ${\bf a}_i$ cycles sit on the $i$-th brane.
The $\mathbb{Z}_2$ involution takes the upper half-plane to the lower half-plane.  Let
\be
A_{jk}= \int_{{\bf a}_k} \omega_{j,\{g\}} ~~~,~~~  B_{jk} = \int_{{\bf b}_k} \omega_{j,\{g\}}\, .
\label{abjk}
\ee
Here the ${\bf b}_k$ cycle is defined as follows.
If $P_k$ is a point on the ${\bf a}_k$ cycle and $P_k'$ is its
$\mathbb{Z}_2$ image on
the ${\bf a}_k^{-1}$ cycle, then the ${\bf b}_k$ cycle is the line from
$P_0$ to $P_k$ which is identified with
$P_k'$ (on a different sheet due to $g_k$ twist) via gluing and then from $P_k'$ to $P_0$ in this
different sheet. Note that ${\bf b}_k$ is not a closed contour when $g_k$ is non-trivial. As a result,
the integrals of closed twisted differentials will depend on the choice of the base
point $P_0$. Let $b_{jk} = \int_{P_0}^{P_k} \omega_{j, \{g\}}$ on
the path indicated in the Fig. 6, then
\be
B_{jk} = b_{jk}-g_k \bar{b}_{jk}\, .
\label{bjk}
\ee
In particular, this implies that $g_k^{-\frac{1}{2}} B_{jk}$ is purely imaginary. As noted above,
when $g_k$ is non-trivial, $B_{jk}$ depends on the choice of the base point $P_0$.

Integrating around trivial cycle $\prod_{i=1}^{h-1} ({\bf a}_i {\bf b}_i {\bf a}_i^{-1} {\bf b}_i^{-1})$
gives the condition
\be
\sum_{i=k}^{h-1} (1-g_k^{-1}) A_{jk} =0\, .
\ee
Since at least one of the $g_i$  is not identity (say $g_{h-1}$),
we can always eliminate one of the
$A_{jk}$'s (say $A_{j (h-1)}$). We can furthermore normalize $\omega_{j,\{g\}}$ so that
\be
A_{jk} = g_j^{\frac{1}{2}}\delta_{jk} ~~~,~~~ {\rm for} ~~~ k=1,...,h-2.
\label{ajk}
\ee
Then the above equation implies
\be
A_{j (h-1)}= -g_{h-1}^{\frac{1}{2}} \frac{\sin(\theta_j)}{\sin(\theta_{h-1})}.
\label{ajk1}
\ee
We will also need the holomorphic differentials $\omega_{j \{g^{-1}\}}$
that are twisted oppositely to
$\omega_{j\{g\}}$. We will denote the corresponding periods by
$A'_{jk}$ and $B'_{jk}$ and $b'_{jk}$.
They satisfy all the above equations with $g_k$ replaced by $g_k^{-1}$.

Given any two closed twisted differentials $\rho$ and $\rho'$ that are twisted by $\{g\}$ and
$\{g^{-1}\}$ respectively, one can evaluate, by using the standard methods for the marking shown in Fig. 6, the following integral:
\be
\int_{\Sigma} \rho \wedge \rho' = \sum_{k=1}^{h-1} [B_k A'_k - A_k B'_k
+\sum_{\ell=1}^k (1-g_k g_{\ell}^{-1})A_{\ell} A'_k ]\, ,
\label{areag}
\ee
where $A_k = \int_{{\bf a}_k} \rho$, $A'_k = \int_{{\bf a}_k} \rho'$, $B_k= \int_{{\bf b}_k} \rho$ and $B'_k
=\int_{{\bf b}_k} \rho'$.

The $(1,1)$ forms $\omega_{j,\{g\}}\wedge \bar{\omega}_{k,\{g\}}$
are untwisted and therefore can be
integrated on the Riemann surface. Using (\ref{areag}) and
the periods (\ref{bjk}), (\ref{ajk}) and
(\ref{ajk1}), we find
\be
\int_{\Sigma} \omega_{j,\{g\}}\wedge \bar{\omega}_{k,\{g\}} = \tau_{jk} - \bar{\tau}_{kj}\, ,
\label{area}
\ee
where
\be
\tau_{jk} = D_{jk} - C_{jk}\label{taujk}
\ee
and $C$ is a purely imaginary symmetric $(h-2)\times (h-2)$ matrix
which does not depend on the world-sheet moduli. It is
given by
\be
C_{jk}= -i \frac{\sin(\theta_k) \sin(\theta_j-\theta_{h-1})}
{\sin(\theta_{h-1})} ~~~, ~~~~~ {\rm{for}}~~~ j \geq k .\label{cjk}
\ee
On the other hand, the matrix $D$ depends on the world-sheet moduli and is given in terms of the
periods $B_{jk}$ as follows:
\begin{eqnarray}
D_{jk} &=& \frac{g_k^{-\frac{1}{2}}}
{2i\sin(\theta_{h-1})} \int_{b_k b_{h-1} b_k^{-1} b_{h-1}^{-1}} \omega_{j,\{g\}}\nonumber \\
&=& g_k^{-\frac{1}{2}}B_{jk} - \frac{\sin(\theta_k)}
{\sin(\theta_{h-1})}g_{h-1}^{-\frac{1}{2}} B_{j (h-1)}\, .
\label{djk}
\end{eqnarray}
It is important to note that while $B_{jk}$ depends on the base point $P_0$,
$D_{jk}$ does not. This is due to the fact that the
cycle ${\bf b}_k {\bf b}_{h-1} {\bf b}_k^{-1} {\bf b}_{h-1}^{-1}$
is a closed contour even when $g_k$ is non-trivial.
Due to Eq.(\ref{bjk}), $D$ is also purely imaginary, however it is in general not symmetric. As
a consequence the right hand side of equation (\ref{area}) is purely imaginary and
\be \tau_{jk} - \bar{\tau}_{kj}= 2(\tau^S)_{jk},\ee where
$\tau^S$ is the symmetric part of $\tau$.

Similarly, we can define the corresponding quantities for oppositely twist\-ed differentials and
the corresponding periods $\tau'_{jk}$ satisfy the same equations with
$g_k$ replaced by $g_k^{-1}$. Finally, we have the relation
\be
0=\int_{\Sigma} \omega_{j,\{g\}}\wedge \omega_{k,\{g^{-1}\}} = \tau_{jk}
- \tau'_{kj}.
\ee
In the untwisted case, a similar equation implies that the period matrix is symmetric, however
in the twisted case it only says that the transposed $\tau$ is equal to $\tau'$.

Using the twisted differentials, one can write the classical solution for $Z$ as
\be
dZ = \sum_{j=1}^{h-2}[L_j \omega_{j,\{g\}} +
\tilde{L}_j\bar{\omega}_{j,\{g^{-1}\}}].
\label{sol}
\ee
Integrals around ${\bf a}_k$ cycles should give the windings $n_k v_k$,
which implies that $\int_{a_k} dZ=n_k v_k$. Furthermore, $\int_{P_0}^{P_k} dZ = (m_k+y_k)
w_k$, due to the boundary conditions (\ref{bc}). This implies
\be
\int_{{\bf b}_k} dZ = (m_k+y_k) (w_k - g_k \bar{w}_k) =
2i(m_k+y_k) g_k^{\frac{1}{2}} \frac{T_2}{|v_k|}\, ,
\ee
where in the second equality we used the fact that $w_k.*v_k = T_2$. Here, $y_k$ are shifted by
$y_h \frac{w_h.*v_i}{T_2}$, due to the fact that we are choosing our coordinates so that the $h$-th
brane lies on the real axis.
We can then solve for $L$ and $\tilde{L}$ as
\begin{eqnarray}
L &=& \frac{N}{2} + \frac{1}{2\tau^S} [(\tau^A+C) N  + 2 M] \\
\tilde{L} &=& \frac{N}{2} - \frac{1}{2\tau^S} [(\tau^A+C) N  + 2 M] ,
\label{sol1}
\end{eqnarray}
where $\tau^S$ and $\tau^A$ are the symmetric and anti-symmetric parts of $\tau$, respectively, and $L$, $\tilde{L}$, $N$ and $M$ are $(h-2)$-dimensional
column vectors whose components are respectively $L_j$, $\tilde{L}_j$,
$N_j=n_j |v_j|$ while $M_j$ is given by
\be
M_j = i\frac{T_2}{|v_j|} [ (m_j+y_j) -  \frac{v_h.*v_j}{v_h.*v_{h-1}}(m_{h-1}+y_{h-1})].
\ee
The classical action is
\begin{eqnarray}
S &=&-i\pi(L^T \tau^S L + \tilde{L}^T \tau^S \tilde{L}) + 2\pi i\sum_{j=1}^{h-2} n_j W_j \nonumber\\
&=& i\pi [ N^T \tau^S N + \{2M+(\tau^A+C)N\}^T (\tau^S)^{-1} \{2M+ (\tau^A+C)N\} ]
\nonumber\\
&& +~2\pi i\sum_{j=1}^{h-2} n_j W_j,
\end{eqnarray}
where $W_j$ is the Wilson line on the world-volume of the $j$-th brane.

Poisson resummation over $m_j$ for $j=1,\dots,h-2$ gives rise to the following instanton
contribution
\begin{eqnarray}
Z_{\rm{inst}}&=&[\,\prod_{j=1}^{h-2} \frac{|v_j|}{T_2}\,]
({\rm{det}} \tau^S)^{\frac{1}{2}}
\sum_{n_j, k_j, m_{h-1}} e^{i\pi(N^T \tau^S N  +\frac{1}{4}K^T\tau^S K)-\pi
K^T \tau^A N}\nonumber\\
&~&\times ~e^{2\pi i \sum_{j=1}^{h-2}( k_j {\tilde{y}}_j +n_j W_j )} \,e^{2\pi i \sum_{j=1}^{h-2}k_j \frac{v_h.*v_j}{v_h.*v_{h-1}} m_{h-1} }
e^{-\pi K^T C N},
\label{poisson}
\end{eqnarray}
where $K$ is a column vector with components $K_j = \frac{|v_j|}{T_2}
k_j$ and ${\tilde{y}}_j$ is the effective Wilson line (\ref{wilsond}) in the Dirichlet
direction. The sum over
$n_j$ ($j=1,\dots,h-2$) is over integers that satisfy the constraint
(\ref{nicons}) while the sum over $k_j$ ($j=1,\dots,h-2$) is over all
integers. Finally, the sum over $m_{h-1}$ ranges from $0$ to $p-1$, where $p$ is the smallest
integer such that $p\frac{v_h.*v_j}{v_h.*v_{h-1}}$ is integer for all
$j$. The latter sum reduces the $k_j$ sums to a sublattice satisfying
the constraint
\be
\sum_{j=1}^{h-2} k_j \frac{v_h.*v_j}{v_h.*v_{h-1}} = {\rm integer}.
\label{kicon}
\ee
This constraint is exactly as the one for the $n_j$ and from the above
discussion it follows that this implies that there exist integers
$k_{a}$, $a=1,\dots,h$, such that $\sum_{a} k_{a} v_{a} =0$.
In fact, $K_{a} = k_{a} *v_{a}/T_2$ is a vector in the dual lattice and
describes the momentum in the
Dirichlet direction for the ${a}$-th brane (i.e.\ in the direction perpendicular
to $v_{a}$).

The phase $e^{-\pi K^T C N}$ is independent of the world-sheet moduli and of the target
space moduli. One can show using the constraints (\ref{nicons}) and the similar one
for $k_j$, that $i K^T C N$ is an integer implying that this phase is $\pm 1$.

We can also include a $B$ field on the plane. The contribution of the $B$ field to the
instanton action is
\be
S_B = \frac{2\pi B}{T_2}  N^T (M+ \frac{1}{2}CN).
\ee
Including this contribution (after the Poisson resummation) in Eq.(\ref{poisson}),
one finally obtains\footnote{Here and in Section 5 [from Eq.(5.18) onwards] we do not keep track of overall factors that are completely moduli- and
flux data-independent.}
\begin{eqnarray}
Z_{\rm{inst}}&=& p\, [\,\prod_{j=1}^{h-2} \frac{|v_j|}{T_2}\,]
({\rm{det}} \tilde{\tau}^S)^{\frac{1}{2}}
\sum_{n_j, k_j } e^{\frac{i\pi}{T_2}(n^T \bar{T} + \frac{1}{2}k^T)
\tilde{\tau} (n T  + \frac{1}{2} k)}\nonumber\\
&~&\times~e^{2\pi i (k_j {\tilde{y}}_j+ n_j W_j)}\, e^{-\pi K^T C N},
\label{zinst}
\end{eqnarray}
where $n$ and $k$ are $(h-2)$-dimensional column vectors with integer entries $n_j$ and $k_j$
satisfying the constraints (\ref{nicons}) and (\ref{kicon}), respectively, and
\be
\tilde{\tau}_{jk} \equiv \frac{|v_j||v_k|}{T_2} \tau_{jk}.
\ee

It is worth noting that $\tau_{jk}$ depend only on the world-sheet moduli and the
twist angles. In turn, these angles depend on the target space modulus $U$, but not on $T$.
Since $\frac{|v_j||v_k|}{T_2}$ does not depend on $T$ as well, it follows that ${\tilde{\tau}}$
also only depends on $U$,  but not on $T$. Thus the $T$-dependence of the partition
function is explicit in (\ref{zinst}). The $U$ dependence, however, is implicit and appears
through ${\tilde{\tau}}$.

Let us now consider the degeneration limit of $\tau$ when ${\bf a}_1$ cycle vanishes. In
Fig. 6, this amounts to shrinking ${\bf a}_1$ and ${\bf a}_1^{-1}$ cycles to points $z_1$ and its
image, respectively. In this limit,
the usual period matrix $t$ given by the untwisted differentials degenerates as $t_{11}
\rightarrow i\infty$. Since the twisted $\omega_1$ will degenerate to a twisted differential
with two simple poles at $z_1$ and its image, with residues $g_1^{\frac{1}{2}}$ and
$-g_1^{-\frac{1}{2}}$, respectively. Then Eqs. (\ref{taujk}), (\ref{cjk}), (\ref{djk})
and (\ref{bjk}) imply that $\tau_{11} \rightarrow
t_{11}$ modulo finite terms. Thus the $t_{11}$ dependence of the partition function is
\be
q_1^{\frac{1}{2} [n_1 \vec{v}_1 + (k_1+B n_1) \frac{*\vec{v}_1}{T_2} ]^2}  ~~~, ~~~~
q_1=e^{2i\pi t_{11}}  .
\ee
This agrees with the fact that the boundary state describing the brane parallel to the vector
$\vec{v}_1$ involves precisely the lattice states
$[n_1 \vec{v}_1 + (k_1+B n_1) \frac{*\vec{v}_1}{T_2}]$. As a further check, one might ask what happens
when we take the limit of shrinking the ${\bf a}_{h-1}$ cycle to a point $z_{h-1}$.
Since in our treatment the $(h-1)$-th
boundary played a special role, it is not immediately obvious that the corresponding degeneration
would have a correct interpretation in terms of the boundary state. In this limit all the basis
elements of twisted differentials $\omega_{j,\{g\}}$ develop simple
poles at $z_{h-1}$ and its image,
with residues $-\frac{\sin(\theta_j)}{\sin(\theta_{h-1})} g_{h-1}^{\pm \frac{1}{2}}$, respectively.
Then Eqs. (\ref{taujk}), (\ref{cjk}), (\ref{djk})
and (\ref{bjk}) imply that the leading behavior of the twisted period matrix is given by
\be
\tau_{jk} \rightarrow \frac{\sin(\theta_j) \sin(\theta_k)}{\sin(\theta_{h-1})^2} t_{(h-1)(h-1)}
\ee
and the $q_{h-1}\equiv t_{(h-1)(h-1)}$ dependence of the partition function is
\begin{eqnarray}
&~&q_{h-1}^{\frac{1}{2}\sum_{j,k=1}^{h-2}
\frac{\sin(\theta_j) \sin(\theta_k)}{\sin(\theta_{h-1})^2}\frac{|v_j||v_k|}{T_2^2}
(n_j T + k_j)(n_k \bar{T}+k_k)}  \\
&~&=q_{h-1}^{\frac{1}{2} [n_{h-1} \vec{v}_{h-1} + (k_{h-1}+B n_{h-1}) \frac{*\vec{v}_{h-1}}{T_2} ]^2},
\end{eqnarray}
where we used the constraints (\ref{nicons}) and (\ref{kicon}) for $n_j$ and $k_j$. This again agrees with
the boundary state of the $(h-1)$-th brane.

Finally, let us consider the case when all the $h-1$ ${\bf a}_j$-cycles are shrunk to points $z_j$ for
$j=1,\dots,h-1$. Then the partition function must describe a $(h-1)$-point function on the disk whose boundary sits on the
$h$-th brane. The vertices at the $h-1$ points are boundary states on the $j$-th brane, namely
\be
V(z_j, n_j, k_j)= e^{i n_j \vec{v}_j.(\vec{X}_L-\vec{X}_R) + \frac{i}{T_2}(k_j+ Bn_j)*\vec{v}_j.
(\vec{X}_L+\vec{X}_R)} (z_j)
\ee
for integers $n_j$ and $k_j$ satisfying the constraints (\ref{nicons}), (\ref{kicon}). To see this, we note that, in this limit,
$\omega_i$ become
\be
\omega_i(z) = \frac{1}{2\pi i} [ \frac{e^{i\theta_i}}{z-z_i} -  \frac{e^{-i\theta_i}}{z-\bar{z}_i}
-\frac{\sin(\theta_i)}{\sin(\theta_{h-1})}( \frac{e^{i\theta_{h-1}}}{z-z_{h-1}} -
 \frac{e^{-i\theta_{h-1}}}{z-\bar{z}_{h-1}} )]\, .
\ee
Using this we can compute $\tau$ and, after some algebra and taking care of the logarithmic
branch cuts, we indeed find that the result is
\be
\sum_{n_j, k_j} \langle\prod_{j=1}^{h-1} q_j^{\Delta(n_j, k_j)} V(z_j, n_j, k_j)\rangle\, .
\ee

So far, we have discussed the contribution of the classical solutions to the partition function.
In the topological amplitude discussed in this paper we actually need the correlation function
\be
\langle\prod_{j=1}^{h-2} [\bar{\psi}\partial Z (w_j)+\bar{\psi}\partial Z (w'_j)]  \rangle_h,
\ee
where $w'_j$
is the image of $w_j$. In the topological theory ${\bar{\psi}}$ are dimension one fields and
have $h-2$ zero modes and therefore are replaced by the $h-2$ twisted differentials $\omega_i$.
Similarly plugging in the classical solution for $\partial Z$ (\ref{sol}), (\ref{sol1})
and doing the Poisson resummation over the integers $m_j$, we find that the world-sheet
supercurrents in the correlation function are replaced by
$\det(\Omega_i (w_j))$, where \be\Omega_i = \sum_{j=1}^{h-2} \frac{1}{T_2}(n_{j} T + k_{j})
v_{j}(\omega_{j,{g}} \omega_{i,{g^{-1}}}+ {\rm c.c.}).\ee

The above discussion has been limited to the case of magnetized D9 branes on $T^6$,
and as a result the boundaries
(${\bf a}$-cycles) are untwisted. This allowed us to normalize the Prym
differentials along the ${\bf a}_i$-cycles for $i=1,\dots, h-2$.
One can generalize the above to the case of orbifolds or fractional branes where some (or all) of the
{\bf a}-cycles are also twisted. The twisted ${\bf a}_i$ cycles are now not closed, however, we can choose instead
the closed cycles ${\bf a}_i {\bf b}_i {\bf a}_i^{-1} {\bf b}_i^{-1}$ to normalize the Prym differentials. This analysis has been carried
out in Ref.\cite{nsv} in the context of closed strings. It is straightforward,
to generalize the method of this reference
to the case of open strings.

\end{document}